\documentclass[fleqn,usenatbib]{mnras}

\usepackage{mathptmx}
\usepackage{amsmath}
\usepackage{amssymb}

\usepackage[T1]{fontenc}
\usepackage{ae,aecompl}
\usepackage{soul} 
\usepackage{sidecap}

\usepackage{graphicx}	
\usepackage{sidecap}
\usepackage{amsmath}	
\usepackage{amssymb}	
\usepackage{color}

\newcommand{\Msun}{M$_{\odot}$}
\newcommand{\Lsun}{L$_{\odot}$}

\newcommand{\mic}{$\mu$m}

\newcommand{\Myr}{M$_{\odot}$\,yr$^{-1}$}
\usepackage{adjustbox}
\usepackage[graphicx]{realboxes}
\usepackage{rotating}

\title[Variability properties of the Arecibo sample]{An infrared study of Galactic OH/IR stars. III. Variability properties of the Arecibo sample}

\author[Jim\'enez-Esteban et al.]{
F. M. Jim\'enez-Esteban,$^{1}$\thanks{E-mail: fran.jimenez-esteban@cab.inta-csic.es}
D. Engels,$^{2}$
D. S. Aguado,$^{3}$
J. B. Gonz\'alez,$^{4}$
P. Garc\'{i}a-Lario$^{5}$
\\
$^{1}$Departamento de Astrof\'{\i}sica, Centro de Astrobiolog\'{\i}a (CSIC-INTA), ESAC Campus, Camino Bajo del Castillo s/n\\
~~E-28692 Villanueva de la Ca\~nada, Madrid, Spain; Spanish Virtual Observatory, Spain\\
$^{2}$Hamburger Sternwarte, Universit\"at Hamburg, Gojenbergsweg 112, D-21029 Hamburg, Germany\\
$^{3}$Institute of Astronomy, University of Cambridge, Madingley Road, Cambridge CB3 0HA, UK \\
$^{4}$MAX IV Laboratory, Lund University, Box 188, SE-221 00, Lund, Sweden\\
$^{5}$European Space Astronomy Centre (ESAC/ESA), Villanueva de la Ca\~nada, E-28692 Madrid, Spain}

\date{Accepted 2021 May 6. Received 2021 May 6; in original form 2021 March 2}

\pubyear{2020}

\begin{document}
\label{firstpage}
\pagerange{\pageref{firstpage}$-$\pageref{lastpage}}
\maketitle

\begin{abstract} 
We present the results of a near-infrared (NIR) monitoring program carried out between 1999 and 2005 to determine the variability properties of the `Arecibo sample of OH/IR stars'. The sample consists of 385 IRAS-selected Asymptotic Giant Branch (AGB) candidates, for which their O-rich chemistry has been proven by the detection of 1612 MHz OH maser emission. The monitoring data were complemented by data collected from public optical and NIR surveys. We fitted the light curves obtained in the optical and NIR bands with a model using an asymmetric cosine function, and derived a period for 345 sources ($\sim 90$\% of the sample). Based on their variability properties, most of the Arecibo sources are classified as long-period large-amplitude variable (LPLAV) stars, 4\% as (candidate) post-AGB stars, and 3\% remain unclassified although they are likely post-AGB stars or highly obscured AGB stars. The period distribution of the LPLAVs peaks at $\sim400$ d, with periods between 300 and 800 d for most of the sources, and has a long tail up to $\sim2100$ d. Typically, the amplitudes are between 1 and 3 mag in the NIR and between 2 and 6 mag in the optical. We find correlations between periods and amplitudes, with larger amplitudes associated with longer periods, as well as between the period and the infrared colours, with the longer periods linked to the redder sources. Among the post-AGB stars, the light curve of IRAS\,19566+3423 was exceptional, showing a large systematic increase (>\,0.4 mag\,yr$^{-1}$) in $K$-band brightness over 7 yr.
\end{abstract}








\section{Introduction}

Low- and intermediate-mass stars (M\,$\lesssim$\,8 \Msun) evolving on the Asymptotic Giant Branch (AGB) are characterized by brightness variability with periods of several hundred days and large amplitudes. Mira variables and a large fraction of the so-called `OH/IR stars' belong to this population of long-period variable (LPV) stars. Their stellar properties are fundamentally determined by pulsations, which are traced by the brightness variations. In period--luminosity diagrams, AGB stars form distinct sequences depending on their primary pulsation mode \citep[][and references therein]{Lebzelter19}. During their evolution on the AGB they develop along these sequences, increasing period and luminosity, and eventually switch to lower radial orders \citep{Trabucchi17}. The switches are apparently connected to the onset of enhanced mass-loss and the pulsations themselves are instrumental for the rates at which the mass-loss occurs \citep{McDonald19}.

The knowledge of the variability properties of AGB stars is therefore key for the derivation of their stellar properties, including the determination of their evolutionary state. The pulsation periods are basic ingredients of scaling relations used to determine mass-loss rates \citep{Goldman17} and luminosities. In particular, the period--luminosity relation for Mira variables for periods P\,$\lesssim$\,500 d \citep{Guandalini08,Whitelock08,Yuan17}, which is attributed to fundamental mode pulsation, plays an important role in determining distances in our Galaxy \citep{Catchpole16,Urago20}. Also, it is seen as a promising additional method to determine distances to nearby galaxies \citep{Whitelock13,Huang18,Huang20,Rau19}.

Stars begin their evolution on the AGB phase with oxygen-rich chemistry. Part of them, those in the main-sequence mass range of 1.5\,$\lesssim$\,M$_{MS}$\,$\lesssim$\,3 \Msun, will convert to carbon-rich chemistry along the AGB because of recurrent dredge up of processed material from the stellar interior. Models adopting solar metallicity predict that stars in this mass range spend the last $4-15$\% of their lifetime on the AGB as carbon rich \citep{DiCriscienzo16}. At the lowest masses (M$_{MS}$\,$\lesssim$\,1.5 \Msun) not enough carbon is dredged up to increase the carbon-to-oxygen ratio on the stellar surface above unity (C/O\,$>$\,1), while for the intermediate-mass stars (M$_{MS}$\,$\ge$\,3.0 \Msun) the `hot-bottom burning' mechanism rapidly converts carbon into nitrogen. So in both the lowest and the highest mass ranges the stars remain oxygen rich until their departure from the AGB.

The mass-loss rates achieved on the AGB are in the range of 10$^{-8}$\,$-$\,10$^{-4}$ \Myr\ \citep[e.g.][and references therein]{DeBeck10}, which lead to the formation of circumstellar envelopes (CSEs) of gas and dust. For the highest mass-loss rates, the CSEs become optically thick at visual light and, for the most extreme cases, at near-infrared (NIR: $1-5$\micron) wavelengths too. Variable OH/IR stars with high mass-loss rates ($\ge$\,$10^{-6}$ \Myr) are examples of such obscured stars in their final stage of AGB evolution. They are predominantly located close to the Galactic plane \citep{Likkel89}, so they are likely descendants of intermediate-mass stars that must have escaped conversion to C-rich chemistry. Optically invisible OH/IR stars were discovered originally by systematic OH maser surveys along the Galactic plane \citep[e.g.][]{Baud81b}, and are named `extreme OH/IR stars' to differentiate them from the more general definition used in the context of the Arecibo sample, which also includes optically bright sources, as for example Mira variables with OH maser emission. 

In addition, the extreme OH/IR star samples discovered by OH maser surveys are a mixed group of variable and non-variable OH/IR stars \citep{Herman85a}. The former show Mira-like large-amplitude variations and are AGB stars. The latter show non-periodic small-amplitude brightness variations, or even no variation, and are assumed to have already evolved off from the AGB. Therefore, they are post-AGB stars still hosting OH maser emission \citep{Engels02}. One way to distinguish between these two kinds of sources is studying their variability properties. 

The AGB evolution models for low metallicity have been tested against the AGB population of the Magellanic Clouds (see \citealt{Pastorelli19}), while for solar metallicity this is hampered by the lack of suitable samples of AGB stars with reliable distances in our Galaxy. Intermediate-mass stars spend most of their time on the AGB at high mass-loss rates \citep{Mouhcine02}, making them underrepresented in AGB samples selected visually or even in the NIR. Long-period (P\,$>$\,1000 d) Mira variables have been discovered in various nearby galaxies by \cite{Karambelkar19}, but in the Magellanic Clouds there seems to be a dearth of counterparts to the Galactic extreme OH/IR stars \citep{Goldman17,Goldman18}. This suggests that perhaps the evolution of the intermediate-mass AGB stars in our Galaxy is observationally not well constrained. Existing samples in the solar neighbourhood (\citealt[][and references therein]{Jura93}; \citealt{Ortiz96,Olivier01}) are incomplete (cf. \citealt{Whitelock94}) and/or biased against stars obscured at wavelengths $\lambda$\,$\le$\,5 \micron. New galactic samples can now be compiled \citep{DiCriscienzo16} using the parallaxes provided by \emph{Gaia} \citep{GaiaCollaboration-Prusti16}; however, they will still exclude the visually obscured stars, which can make up to $\sim40$\% (L\'opez Mart\'i et al., private communication) of the total number of catalogued AGB stars with OH masers \citep{Engels15b}, and $\sim33$\% of the AGB star candidates in the BAaDE survey \citep{Quiroga20}.

Because of the long periods involved ($\sim1-5$ yr), the determination of the variability properties of AGB stars is time consuming and was, before the start of automated optical surveys such as ASAS-SN \citep{Kochanek17} and \emph{Gaia} DR2 \citep{GaiaCollaboration18-Brown}, a domain of amateur observers working in the visual spectral range \citep{Samus17,Kafka19}, or the result of several monitoring programs in the NIR. Galactic AGB star samples with known periods are therefore biased towards visually and NIR bright stars, with only a few monitoring programs performed in the NIR \citep[e.g.][]{Jimenez-Esteban06b,Tang08,Urago20} covering the visually obscured stars, and a few others performed at radio wavelengths \citep[e.g.][]{Herman85a,Engels19} covering the NIR obscured stars.

The Arecibo survey of OH/IR stars \citep{Lewis94} is exceptional, in the sense that it represents a sample of O-rich AGB stars selected in the mid-infrared containing both types, bright and obscured stars at $\lambda$\,$<$\,5 \micron. Because of its well documented selection criteria it offers the possibility to make a new census of the AGB population in the solar neighbourhood. To determine its variability properties, we monitored a large fraction of the sample in the NIR between 1999 and 2005, excluding only visually bright sources already covered by optical monitoring programs. Our initial observations were described in \cite{Jimenez-Esteban05} (hereafter Paper\,I). In \cite{Jimenez-Esteban06a}, we extended the Arecibo sample with highly obscured OH/IR stars taken from the GLMP catalogue \citep{Garcia-Lario92}. This paper continues these multi-epoch analyses with the determination of the variability properties of the Arecibo sample.

The paper is structured as follows. In Section \ref{sample}, we present the Arecibo sample of OH/IR stars. In Section \ref{sec:NIR-MP}, we describe the results of our NIR monitoring program (NIR-MP), and in Section \ref{sec:optical} we describe the set of additional optical light curves obtained with data taken from various public surveys. Section \ref {sec:LC-fit} presents the light-curve analysis and the methods used for the derivation of amplitudes and periods. We analyze the variability properties of the sample in Section \ref{sec:discussion}, and classify the individual sources accordingly. The main conclusions are presented in Section \ref{sec:conclusions}.


\section{The Arecibo sample}
\label{sample} 

The Arecibo sample of OH/IR stars is a well-defined sample of IRAS sources with appropriate colours of AGB stars and 1612 MHz OH maser line detection with the Arecibo radio telescope. It was obtained from a complete survey of IRAS Point Source Catalogue (PSC) \citep{Beichman88} sources with flux densities $\ge$\,2 Jy at 25 \mic\ and declination 0\degr\,$<$\,$\delta$\,$<$\,37\degr\ \citep{Lewis94}. The sample varied in size over the years, because of re-classification of the OH masers and new detections during revisits of the initial IRAS sample. The Arecibo sample as used in this paper was originally defined by \cite{Engels96} containing 389 IRAS sources. Four sources were removed from the sample in Paper\,I because three were identify as molecular clouds and one was an erroneous entry in the IRAS-PSC. Thus, the sample presented in this paper is made of N\,=\,385 IRAS sources. We used the AllWISE catalogue \citep{Wright10} to obtain accurate coordinates for the Arecibo sources. Only in the case that AllWISE coordinates were not available, we used the 2MASS-PSC \citep{2MASS-PSC} instead. The typical deviation from the coordinates given in Paper\,I is smaller than 1\arcsec, but the coordinates of individual sources may deviate by up to 6\arcsec\ in the most extreme cases. Table\,\ref{t:sample}
contains the full list of sources, their coordinates with the reference from which this information has been extracted, and their classification as derived in Sect\,\ref{sec:classification}. The full table is available online (see Data Availability).


\begin{table*}
\caption[]{The Arecibo sample of OH/IR stars. The full table is available electronically at the CDS. Column `Ref' gives the reference from which the coordinates (AllWISE or 2MASS catalogues) have been extracted, and column `Group' contains the assigned membership to one of the four groups (1: variable AGB stars; 2: post-AGB stars; 3: unclassified; 4: miscellaneous sources) used in our classification scheme (see Sect\,\ref{sec:classification}).}
\label{t:sample}
\begin{center}
\begin{tabular}[t]{llcccccc}
\hline\hline\noalign{\smallskip}
\multicolumn{2}{c}{Names} &\,\,\,& \multicolumn{3}{c}{Coordinates (2000)}  &\,\,\,& Group\\
\cline{1-2} \cline{4-6} 
\noalign{\smallskip}
IRAS      &     Other    &&  RA & Dec. & Ref && \\
\hline\noalign{\smallskip}
  01037+1219 & IRC +10011 && 01:06:26.02 & +12:35:53.2 & AllWISE && 1\\
  01085+3022 & IRC +30021 && 01:11:15.95 & +30:38:06.5 & AllWISE && 1\\
  02404+2150 & YY Ari     && 02:43:16.48 & +22:03:34.9 & AllWISE && 4\\
  02420+1206 & RU Ari     && 02:44:45.53 & +12:19:03.2 & AllWISE && 1\\
  02547+1106 & YZ Ari     && 02:57:27.48 & +11:18:05.8 & AllWISE && 1\\
  03507+1115 & IK Tau     && 03:53:28.87 & +11:24:21.7 & 2MASS   && 1\\
\noalign{\smallskip}\hline
\end{tabular} 
\end{center}
\end{table*}

\begin{figure}
\includegraphics[width=50mm,trim={ 0 0 4.0cm 1cm}, angle=90]{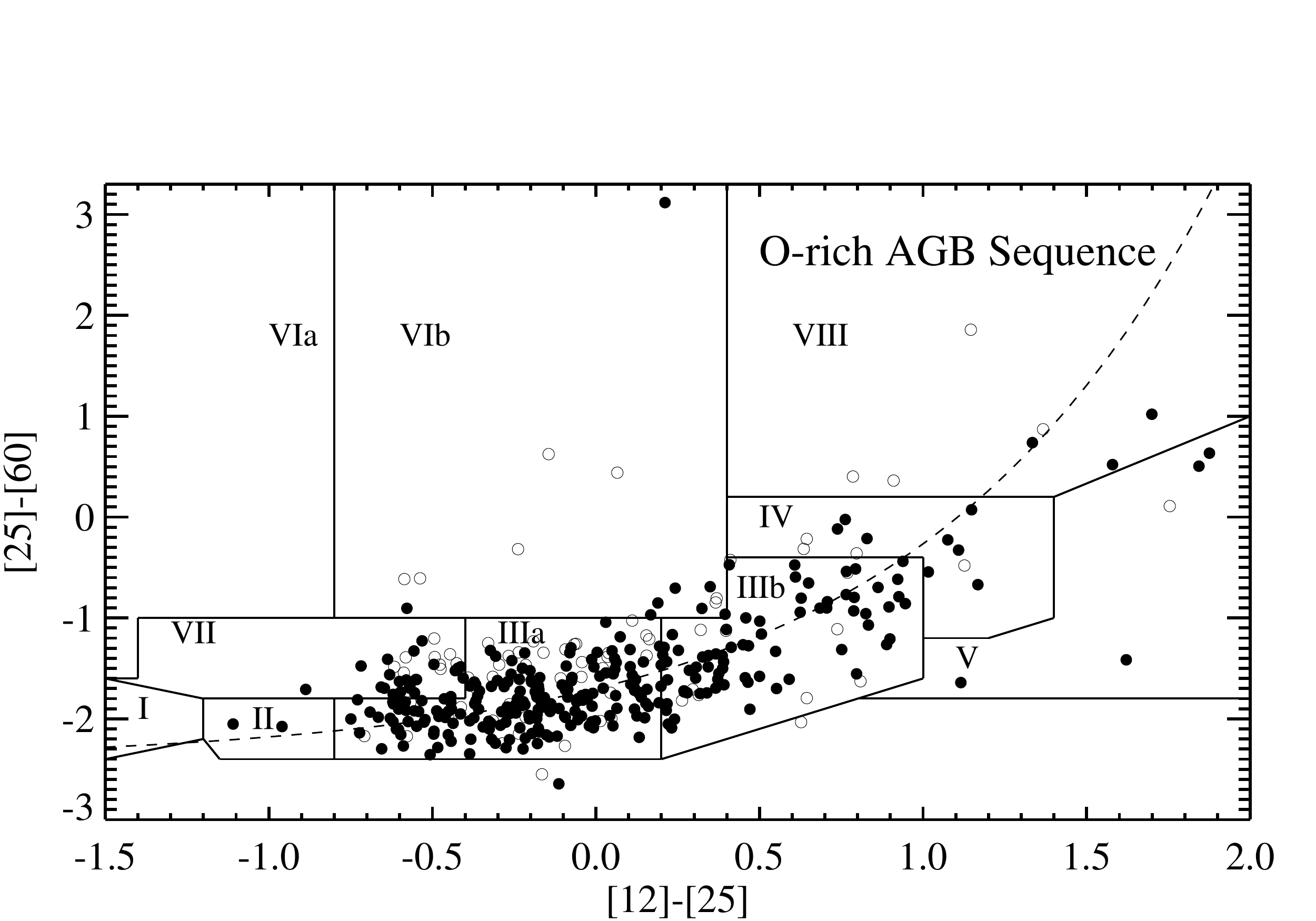}
  \caption{The position of the Arecibo sample in the IRAS 2CD. The IRAS colours are defined as in Paper\,I and are obtained from the IRAS-PSC, version 2.0. The original selection of the sample was made in an earlier version of the IRAS-PSC. Those sources marked with open symbols have now only an upper limit in one of the IRAS filters. The dashed line is the `oxygen-rich AGB sequence' (see the text). The boxes were defined by \citet{vanderVeen88a}.}
  \label{fig:IRAScc}
\end{figure}

In Fig.\,\ref{fig:IRAScc}, we show the position of the sources of the Arecibo sample in the IRAS two-colour diagram (2CD). Most sources are located in the boxes IIIa and IIIb, which were introduced by \cite{vanderVeen88a} to describe the location of O-rich AGB stars. The dashed line represents the sequence of IRAS colours expected for oxygen-rich AGB stars with increasing mass-loss rate \citep{Bedijn87}, hereafter the `O-rich AGB sequence'. 

Paper\,I studied the optical and NIR properties of the Arecibo sample. Approximately 2/3 of them showed an optical counterpart in the red filter (6700 \AA) of the Second Digitized Sky Survey (\citealt{Djorgovski01}; detection limit of $\sim20.8$ mag), and therefore most likely have an optically thin CSE, consistent with their location in the IRAS 2CD. The Arecibo sample predominantly represents the bluer population of OH/IR stars, but also contains a large fraction ($\sim$\,1/3) of visually obscured extreme OH/IR stars.


\section{Near infrared monitoring program}
\label{sec:NIR-MP}

The near-infrared monitoring program (NIR-MP) described in this paper was carried out over a period of 6 yr, from 1999 June to 2005 June. It entails 18 observing runs, adding up 160 nights with four different telescopes (see below). All 385 sources of the Arecibo sample were observed in the three broad-band filters $J$ (1.25\,\mic), $H$ (1.65\,\mic), and $K$ (2.2\,\mic), with the only exception of the very bright source IRAS\,19039+0809 (R Aql) because of saturation issues. However, not all sources were observed with the same frequency. At the beginning of the NIR-MP in 1999, we assigned a lower priority to those sources for which variability information was already available in the optical (a small part of the sample), mostly those sources with the brightest optical counterparts. They were only observed as fillers at the end of every run after observing the rest of (less studied) sources. Moreover, depending on the colour of the sources the monitoring was also not performed necessarily in all three filters, but only in those where detection was expected.

\subsection{Observations} 

The NIR-MP was carried out using the 1.23, the 2.2, and 1.52 m EOCA\footnote{\url{http://astronomia.ign.es/web/guest/instalaciones\#EOCA}} telescopes at the Calar Alto Observatory (CAHA\footnote{\url{https://www.caha.es/}}, Almer\'{i}a, Spain), always equipped with the MAGIC NIR camera \citep{Herbst93}, and the 1.5 m Carlos S\'anchez Telescope (CST) equipped with the CAIN NIR camera at Teide Observatory (TO\footnote{\url{https://www.iac.es/es/observatorios-de-canarias/observatorio-del-teide}}, Islas Canarias, Spain). The individual observing runs are listed in Table\,\ref{t:obs-run}. 

\begin{table}
\caption[]
{\label{t:obs-run}Observing runs.}
\begin{center}
\begin{tabular}[]{lllr}
\hline\hline\noalign{\smallskip}
Run & Telescope & Dates & Nights \\ 
\hline\noalign{\smallskip}
\,\,\,1 & 1.23\,m & 1999 Jun 30 to Jul 09 & 10 \\
\,\,\,2 & 1.23\,m & 2000 Jul 18 to Jul 27 & 10 \\
\,\,\,3 & 1.23\,m & 2001 May 04 to May 16 & 13 \\
\,\,\,4 & 1.23\,m & 2001 Jul 27 to Aug 03 &  8 \\
\,\,\,5 & 1.23\,m & 2001 Sep 07 to Sep 09 &  3 \\
\,\,\,6 & 1.23\,m & 2002 May 17 to Jun 09 & 24 \\
\,\,\,7 & 1.23\,m & 2002 Jul 29 to Aug 09 & 12 \\
\,\,\,8 & 1.23\,m & 2002 Sep 09 to Sep 14 &  6 \\
\,\,\,9$^{a}$ & CST & 2003 May 07, Jul 14 and 25, Aug 14 &  4 \\
10 &  2.2\,m & 2003 Jul 09 to Jul 17 &  9 \\
11 & EOCA    & 2003 Jul 29 to Aug 04 &  7 \\
12 & CST     & 2003 Sep 01 to Sep 09 &  9 \\
13 & EOCA    & 2004 May 04 to May 10 &  6 \\
14 & 2.2\,m  & 2004 Jun 26 to Jun 30 &  5 \\
15 & EOCA    & 2004 Jul 06 to Jul 14 &  9 \\
16 & CST     & 2004 Aug 06 to Aug 09 &  4 \\
17 & CST     & 2004 Sep 13 to Sep 22 & 10 \\
18 & EOCA    & 2005 Jun 08 to Jun 18 & 11\\
\noalign{\smallskip}\hline
\end{tabular} 
\end{center}
$^{a}$Service mode.
\end{table}

The MAGIC camera had a 256\,$\times$\,256 NICMOS3 HgCdTe detector array operating with pixel scales of 1.2, 0.6, and 0.9 arcsec, and an approximate field of view of 5\arcmin\,$\times$\,5\arcmin, 2.7\arcmin\,$\times$\,2.7\arcmin, and 4\arcmin\,$\times$\,4\arcmin, for the 1.23\,m, 2.2\,m, and EOCA telescopes at CAHA, respectively. Similar to the MAGIC camera, the CAIN camera installed at the 1.5 m CST telescope at TO, also used a 256\,$\times$\,256 NICMOS3 HgCdTe detector array with a pixel scale of 1\arcsec\ in this case, which provided an approximate field of view of 4.3\arcmin\,$\times$\,4.3\arcmin.

To calibrate the images obtained, we followed a twofold strategy. Generally, relative photometry was planned using as reference the photometry of field stars from the 2MASS-PSC \citep{2MASS-PSC}. However, in order to avoid saturation only very short exposure times were possible for the brightest sources. This resulted in very few detected field stars, which made relative photometry either inaccurate or even impossible. These bright sources were identified at the beginning of the monitoring program, and were observed always under atmospheric conditions good enough for absolute photometric calibration. During these nights, standard stars from \cite{Elias82} were observed several times at different airmasses, to obtain the extinction curve for the corresponding night. This strategy allowed us to make the most of the observing time, avoiding to spend time observing standard stars regularly and obtaining reliable relative photometry also under non-photometric atmospheric conditions. 

In order to derive an accurate determination of the background, something critical to achieve a good quality photometry, we used a standard dithering technique, obtaining five images per filter with the source located at different positions in the image. Throughout the whole monitoring program, we collected a total of $\approx$\,150,000 images, considering both calibration and science exposures. A semi-automated procedure was developed to reduce them. We refer the reader to Paper\,I for a more extended explanation of the observation technique and the image data reduction process.

\subsection{Photometry}

We first used the software {\sc Astrometry.net}\footnote{\url{http://astrometry.net}} to perform an astrometric calibration of all scientific images. This allowed us to use the 2MASS catalogue for relative photometry at a later step.

Next, we used {\sc SExtractor} \citep{Bertin96} on each image to derive aperture photometry of all sources with signal-to-noise 3$\sigma$ above the background. Each photometric measurement consisted of summing up the instrumental counts within the aperture. The measurements obtained from the five individual images of the dithering process were averaged, weighting each individual measurement with its statistical error. The result is an instrumental magnitude which error is the standard deviation of the individual measurement.

Sometimes problems occurred, either in the acquisition or in the data reduction process that yielded to bad individual measurements. To remove these erroneous measurements, we applied up to two times a 3$\sigma$ clipping cut. The fraction of instrumental magnitudes calculated with five individual measurements was $\sim60$\%, while for the rest a minimum of three individual measurements were imposed.

Finally, we performed the photometric calibration of the observations. For the majority of the sample, relative photometry was possible. To obtain the zero point for each observation we fitted a linear relation between our instrumental magnitudes and the 2MASS magnitudes of the field stars. In order to correct for deviating instrumental magnitudes and to discard variable stars in the field, we applied iteratively a 3$\sigma$ clipping process. This allowed us to derive high-quality and reliable photometry. 

For bright source images, where relative photometry was not possible, we carried out an absolute photometric calibration, using measurements of standard stars taken at different airmasses to derive the zero point of the photometric system and the atmospheric extinction. The uncertainties of the photometry were obtained through error propagation throughout the whole process.

\begin{figure}
 \includegraphics[width=\linewidth]{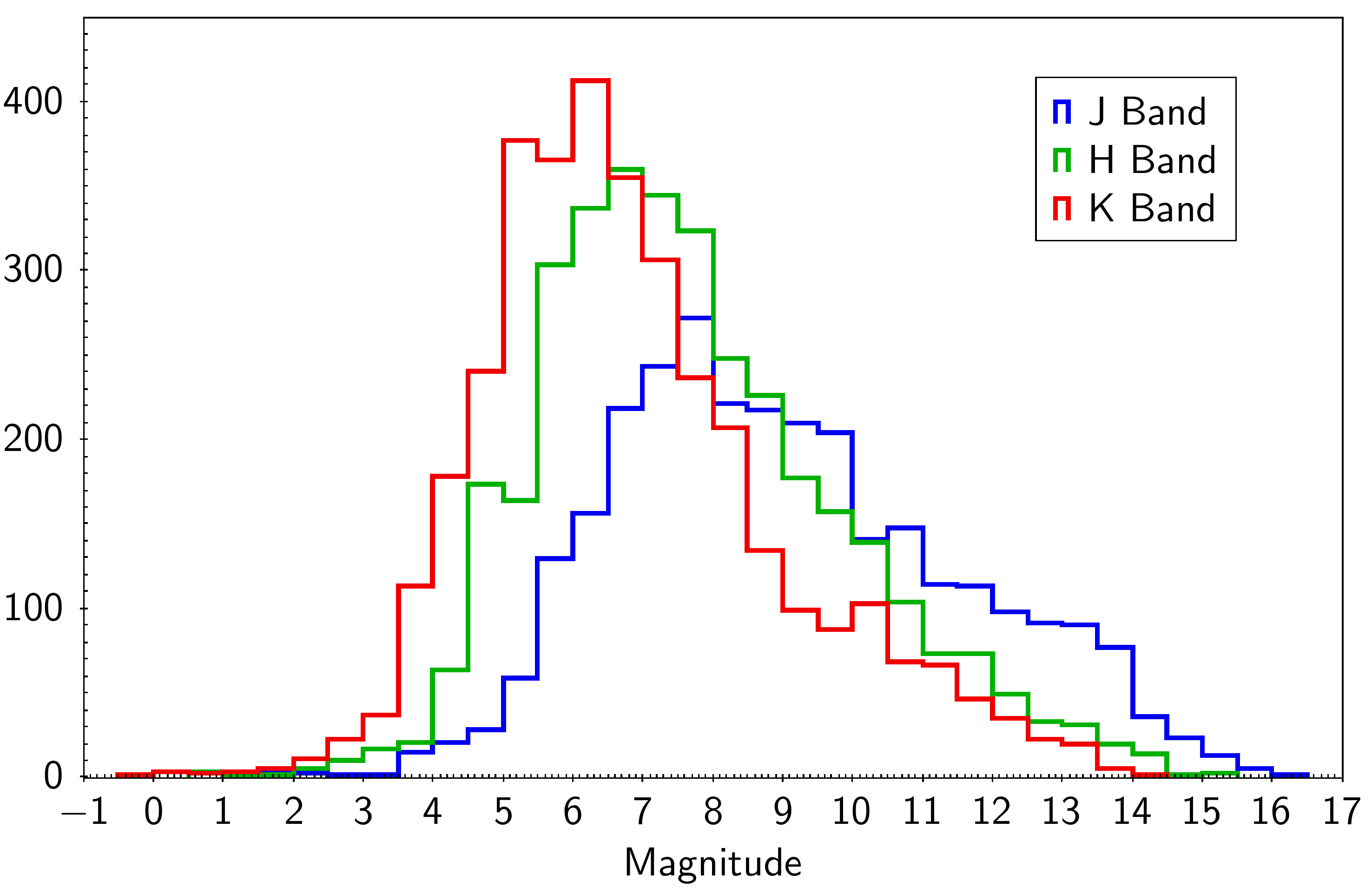}
  \caption{NIR $J$- (blue), $H$- (green), and $K$-band (red) magnitude distributions for the N\,=\,367 Arecibo sources with detections in at least one band in the NIR-MP.}
  \label{F.JHK}
\end{figure}

Our NIR-MP provided $\sim$\,10,000 NIR photometric measurements, with detections having S/N\,$>3\sigma$ for 367 objects of our sample, distributed in the three bands. The number of photometric measurements per source and filter is between 0 and 17. Fig.\,\ref{F.JHK} shows the brightness distribution of the sources in the Arecibo sample. The brightnesses have typical magnitudes around 8, 7, and 6 for the J, H, and K bands, respectively. The sensitivity limits during different observing runs varied with the telescope used and also depended on weather conditions. Typical limits are 16.0, 14.5, and 14.0 mag for J, H, and K, respectively (Fig. \ref{F.JHK}). This is about 1 mag deeper than those achieved during our initial observations reported in Paper\,I. 

We complemented our observational data with data from the public NIR surveys 2MASS, DENIS \citep{DENIS}, and UKIDSS\footnote{\url{http://www.ukidss.org/}} \citep{UKIDSS}. This added $\sim$\,1.200 photometric points to the light curves. Since the overwhelming majority of these sources have large amplitude brightness variations, we did not correct for the small zero-point differences between the K and K$_{s}$ bands used by 2MASS and DENIS. Each individual photometric measurement is listed in Table\,\ref{t:phot}, which is available online (see Data Availability). It contains the observing epoch, the magnitude with its error for each NIR band, and the origin of the data, either from our NIR-MP or from the 2MASS, DENIS, or UKIDSS archival data. The photometry compiled in Table\,\ref{t:phot} is referred to as the NIR data base (NIR-DB), hereafter.

\begin{table}
\caption[]
{\label{t:phot}NIR photometry data for IRAS\,01037+1219. The full table including all Arecibo sources with NIR photometry is available electronically at the CDS. Column `Origin' refers to the source (our NIR-MP, or from 2MASS, DENIS, or UKIDSS surveys).}
\begin{center}
\begin{tabular}[]{lllcl}
\hline\hline\noalign{\smallskip}
  \multicolumn{1}{|c|}{IRAS} &
  \multicolumn{1}{c|}{JD} &
  \multicolumn{1}{c|}{Filter} &
  \multicolumn{1}{c|}{Magnitude} &
  \multicolumn{1}{c|}{Origin} \\
\hline\noalign{\smallskip}
  01037+1219 & 2450697 & $J$ & 7.44$\pm$0.03 & 2MASS\\
  01037+1219 & 2450697 & $H$ & 4.6$\pm$0.2 & 2MASS\\
  01037+1219 & 2450697 & $K$ & 2.2$\pm$0.3 & 2MASS\\
  01037+1219 & 2451747 & $J$ & 8.477$\pm$0.012 & NIR-MP\\
  01037+1219 & 2451747 & $H$ & 5.153$\pm$0.007 & NIR-MP\\
  01037+1219 & 2451747 & $K$ & 3.41$\pm$0.06 & NIR-MP\\
  01037+1219 & 2452490 & $J$ & 9.075$\pm$0.015 & NIR-MP\\
  01037+1219 & 2453195 & $J$ & 9.203$\pm$0.011 & NIR-MP\\
  01037+1219 & 2453195 & $H$ & 5.82$\pm$0.15 & NIR-MP\\
  01037+1219 & 2453195 & $K$ & 3.008$\pm$0.012 & NIR-MP\\
\noalign{\smallskip}\hline
\end{tabular} 
\end{center}
\end{table}


For 18 sources, we got no photometric data in the NIR-MP. These sources are listed in Table\,\ref{t:no-phot} with their classification given in Section \ref{sec:classification}, and the reason for the lack of measurements. Several of them were detected in our NIR-MP, but were so bright that they saturated the detectors (N\,=\,3) or were blended with a field star of similar brightness (N\,=\,2). The rest (N\,=\,13) were below our sensitivity limits, but were detected by the 2MASS and/or UKIDSS surveys, usually with fainter magnitudes.

\begin{table}
\caption[]
{\label{t:no-phot} Arecibo sources without photometric data in the NIR-MP. The classification and group membership are described in Sect\,\ref{sec:classification}}
\begin{center}
\begin{tabular}[]{lcll}
\hline\hline\noalign{\smallskip}
IRAS         & Group  & Classification& Comment \\
\hline\noalign{\smallskip}
05284+1945   & 3 & Unclassified  & Not detected \\
06500+0829   & 1 & LPLAV	     & Saturated \\
09425+3444   & 1 & LPLAV	     & Saturated \\
18475+0353   & 3 & Unclassified  & Not detected \\
18498$-$0017 & 1 & LPLAV	     & Not detected \\
18501+0013   & 3 & Unclassified  & Not detected \\
18517+0037   & 3 & Unclassified  & Not detected \\
18596+0315   & 2 & Post-AGB star & Not detected \\
19006+0624   & 3 & Unclassified  & Not detected \\
19029+0933   & 3 & Unclassified  & Blended \\
19039+0809   & 1 & LPLAV	     & Saturated \\
19067+0811   & 1 & LPLAV	     & Not detected \\
19178+1206   & 3 & Unclassified  & Not detected \\
19188+1057   & 3 & Unclassified  & Not detected \\
19254+1631   & 2 & Post-AGB star & Not detected \\
19374+1626   & 3 & Unclassified  & Not detected \\
19440+2251   & 3 & Unclassified  & Not detected \\
20149+3440   & 4 & YSO           & Blended \\
\noalign{\smallskip}\hline
\end{tabular} 
\end{center}
\end{table}

In Table\,\ref{t:photosummary}, also available online, we present for each object the number of photometric points (N$_{i}$), the amplitude ($\Delta$i) as the difference between the maximum and minimum observed magnitudes, and the arithmetic mean magnitude (<$i$>) of the observational data, where $i$ makes reference to the data band ($J$, $H$, and $K$ for the NIR, and $O$ for the optical). In the last column, we list the origin of the optical data collected from the public surveys (see Section \ref{sec:optical}).

\begin{figure}
\includegraphics[width=\linewidth]{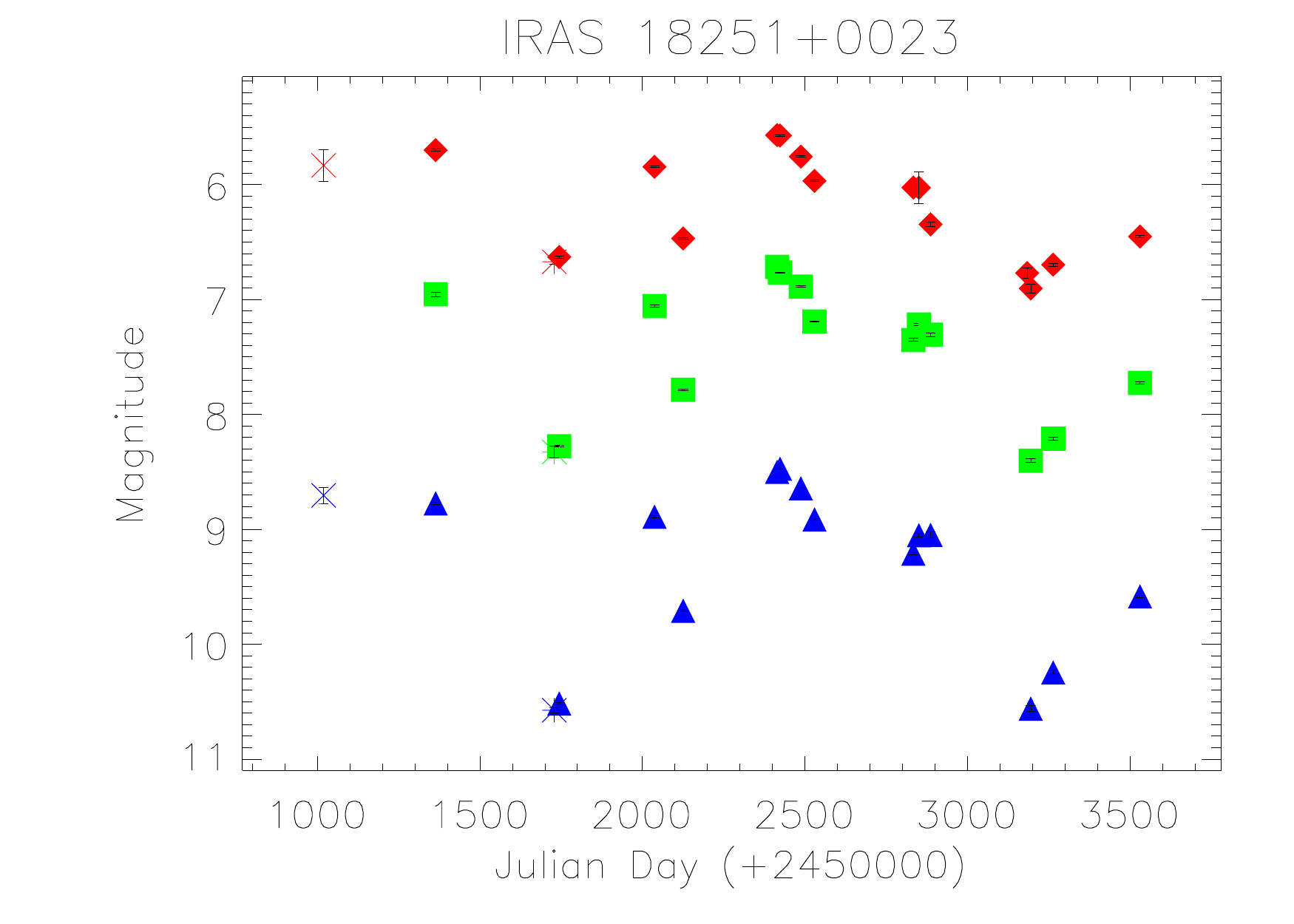}
  \caption{Light curves of IRAS\,18251+0023 in the $J$ (blue triangles), $H$ (green squares), and $K$ (red diamonds) band from the NIR-MP. Additional photometric points are from 2MASS (asterisks) and DENIS (crosses), with the same colour code as the NIR-MP data. Photometric error bars are overplotted.}
  \label{F.18251LC}
\end{figure}

Figure\,\ref{F.18251LC} shows an example of one of the best sampled light curves extracted from the NIR-DB. A representation of each NIR light curve can be found in Appendix\,\ref{Ap.NIR-LC}.

\begin{table*}
\begin{center}
\renewcommand{\tabcolsep}{2pt}
\centering
\caption{Average brightness and peak-to-peak amplitude of the Arecibo sources derived directly from the observational data. The $J$-, $H$-, and $K$-band data are from the NIR-DB, and the optical data are from a combination of data from public surveys. N$_{i}$ gives the number of photometric points, and the last column lists the public surveys used (for details see Section \ref{sec:optical}). The full table is available electronically.}
\label{t:photosummary}
\begin{tabular}{ccrcccrcccrcccrccl}
\hline\hline\noalign{\smallskip}
&\,\,\,&\multicolumn{3}{c}{$J$ band} &\,\,\,& \multicolumn{3}{c}{$H$ band} &\,\,\,& \multicolumn{3}{c}{$K$ band} &\,\,\,& \multicolumn{4}{c}{Optical}\\
\cline{3-5} \cline{7-9} \cline{11-13} \cline{15-18}
\noalign{\smallskip}
IRAS && N$_{J}$ & $\Delta$$J$ & <$J$> && N$_{H}$ & $\Delta$$H$ & <$H$> && N$_{K}$ & $\Delta$$K$ & <$K$> && N$_{O}$ & $\Delta$$O$ & <$O$> & \multicolumn{1}{}{Survey(s)}\\
     &&  & (mag) & (mag) &&  & (mag) & (mag) &&   & (mag) & (mag) &&   & (mag) & (mag) & \\
\hline\noalign{\smallskip}
01037+1219 && 4 & 1.8 & 8.5 && 3 & 1.2 & 5.2 && 3 & 1.2 & 2.9 &&  -- & -- & -- & -- \\
01085+3022 && 8 & 2.0 & 4.9 && 7 & 1.6 & 3.4 && 7 & 1.3 & 2.5 &&  -- & -- & -- & -- \\
02404+2150 && 7 & 0.5 & 5.1 && 6 & 0.6 & 4.2 && 5 & 0.6 & 3.7 &&  200 & 0.8 & 13.7& ASAS-SN\\
02420+1206 && 3 & 1.0 & 4.4 && 3 & 1.4 & 3.7 && 3 & 0.8 & 2.9 && 1216 & 6.5 & 12.8& ASAS and ASAS-SN and AAVSO\\
\noalign{\smallskip}\hline
\end{tabular}
\end{center}
\end{table*}

\section{Optical light curves}
\label{sec:optical}

As stated above, about 2/3 of the Arecibo sample shows an optical counterpart in the Second Digitized Sky Survey. Recently, several optical surveys have been carried out, which provide time series of photometric data. We used them to create optical light curves of as many Arecibo sources as possible. The goal is twofold: (i) From the analysis of the optical light curves (see Section \ref{sec:LC-fit}), we will have an independent determination of the variability periods, which can be used for internal validation of the periods derived from the NIR-DB; and (ii) we can determine the period for those sources for which we could not derive it from the NIR-DB.

The data from the following surveys were used: the \textit{All Sky Automated Survey} (ASAS)\footnote{\url{http://www.astrouw.edu.pl/asas/}} \citep{Pojmanski02}, the \textit{All-Sky Automated Survey for Supernovae} (ASAS-SN)\footnote{\url{https://asas-sn.osu.edu/}} \citep{Kochanek17}, the \textit{American Association of Variable Star Observers} (AAVSO)\footnote{\url{https://www.aavso.org/}} \citep{Kafka19}, the \textit{Optical Monitoring Camera} (OMC) onboard of the \textit{International Gamma Ray Astrophysics Laboratory} (INTEGRAL)\footnote{\url{http://sci.esa.int/integral/}} \citep{Alfonso-Garzon12}, and the \emph{Gaia DR2}\footnote{\url{http://gea.esac.esa.int/archive/}} \citep{GaiaCollaboration18-Brown}.

We used only data with good quality and quoted errors lower than 0.2 mag. For several Arecibo sources, more than one survey provides photometric measurements at similar epochs, just separated by few days. Comparing observations from different surveys made within 5 d, we estimated that the typical brightness differences between surveys were similar to their errors. Consequently, since most of the Arecibo sources were expected to be large-amplitudes variables with changes of several magnitudes in the optical, we did not attempt to correct for the small systematic magnitude offsets between the different photometric systems. 

In the case of the \emph{Gaia} data, we used the $G_{BP}$ band to combine with the $V$-band data from the different surveys, because among the three \emph{Gaia} bands it is the most similar to the V band. Otherwise, when optical data were only available from \emph{Gaia}, we used the $G$-band photometry instead, because more measurements are available and with smaller photometric errors than those in the $G_{BP}$ band.

\begin{figure}
\includegraphics[width=\linewidth]{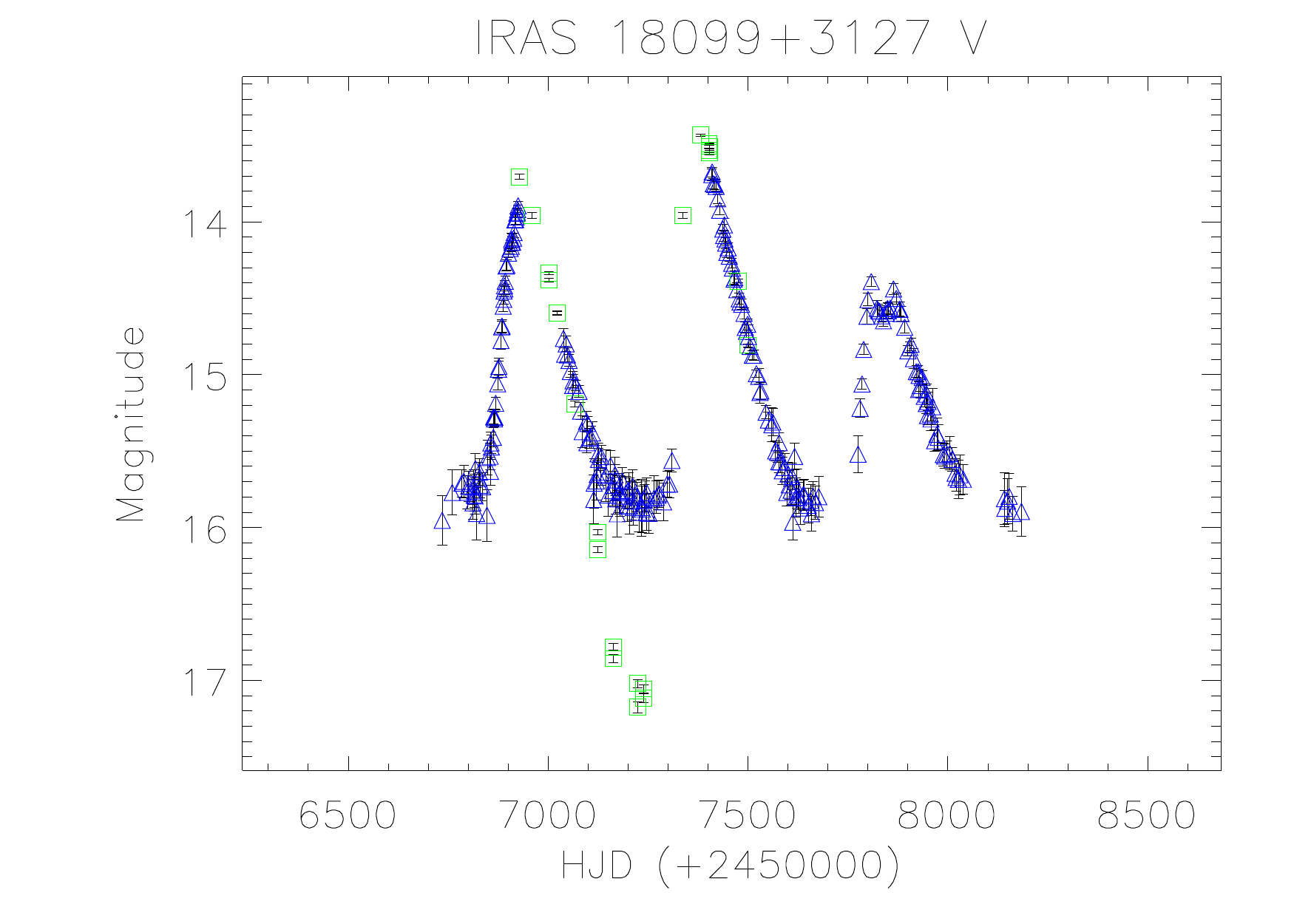}
\includegraphics[width=\linewidth]{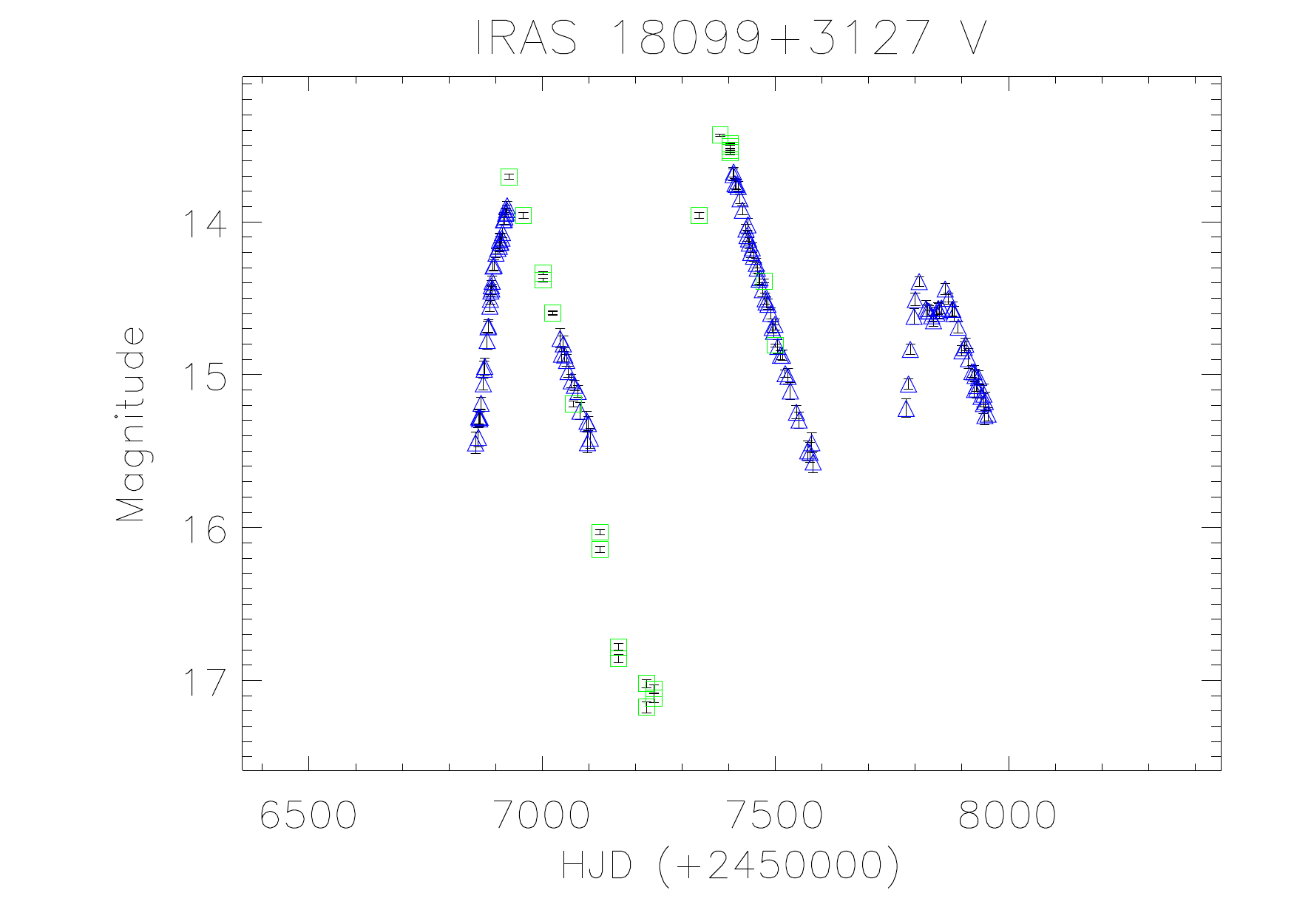}
  \caption{Comparison of the optical $V$-band light curve of IRAS\,18099+3127 before (upper panel) and after (lower panel) the removal of suspected contaminated data from ASAS-SN (blue triangles). The \emph{Gaia} $G_{BP}$ data are shown as green squares.}
  \label{F.18099}
\end{figure}

Finally, we visually inspected all optical light curves and discarded any obvious outliers. This case-by-case cleaning was specially important for a handful of sources with the ASAS-SN data close to the limiting magnitude of the survey. An example is shown in Fig.\,\ref{F.18099}. During the minimum of its light curve, IRAS\,18099+3127 is fainter than 17th mag as shown by the \emph{Gaia} data. However, ASAS-SN provides photometric measurements during this phase with values around 16th mag. As the ASAS-SN observations have a resolution of FWHM\,=\,16\arcsec, we suspect that either background emission or emission from a close field star contaminated the measurements.

This way, we created up to 123 optical light curves, 46 of them with only $G$-band \emph{Gaia} photometry. The optical light curves are generally better sampled than the NIR-DB ones. Only 12\% of the light curves have less than 20 observations, with a minimum of 13 for the light curves obtained solely from the \emph{Gaia} data. About 50\% of the light curves have between 20 and 150 observations, and $\sim38$\% have more than 150 observations. The best sampled light curve is the one for IRAS\,19039+0809 (R Aql) with more than 3500 data points, corresponding to more than 18 yr of monitoring (see Fig.\,\ref{F.19039LC}). All the optical light curves are shown in Appendix\,\ref{Ap.Opt-LC}.

\begin{figure*}
\includegraphics[width=\linewidth]{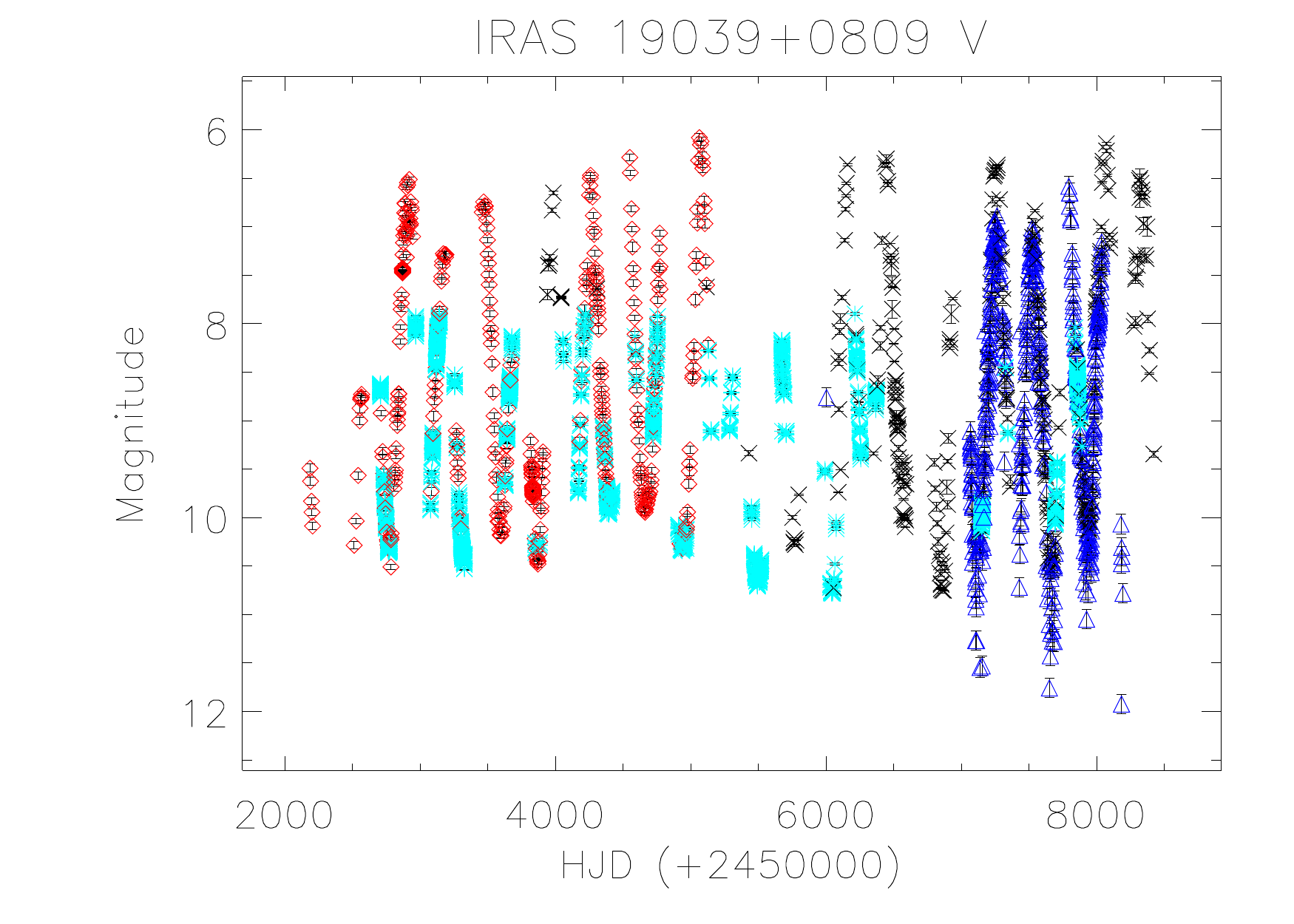}
  \caption{Optical light curve of IRAS\,19039+0809 built with data from ASAS (red diamonds), AAVSO (black crosses), OMC (cyan asterisks), and ASAS-SN (blue triangles). Photometric error bars are shown. They are typically smaller than the symbols.}
  \label{F.19039LC}
\end{figure*}

As for the NIR data, we present in Table\,\ref{t:photosummary} the number of optical photometric points collected, the observed amplitude, derived as the difference between the maximum and minimum observed magnitudes, the arithmetic mean magnitude of the observational data, and the origin of the optical data.


\section{Light-curve analysis}
\label{sec:LC-fit}

\subsection{Modelling of the light curves}

To characterize the variability properties of the LPVs in our sample, we fitted the observational data to a model made of an asymmetric cosine function, characterized by an asymmetry factor and a constant mean magnitude, as in \cite{Jimenez-Esteban06b}. This model is described by the following equations:
\begin{equation}
    m \left( t_i \right)\,=\,\overline{m} + \frac{A}{2} cos\left(2\pi\Omega\left(t_i\right)\right)
\end{equation}
where $m$ is the observed magnitude, $t_i$ is the Julian Date of the observation, $\overline{m}$ is the mean magnitude, $A$ is the peak-to-peak amplitude, and $\Omega\left(t_i\right)$ is: 
\begin{equation}
    \Omega\left(t_i\right)\,=\,\left \{ \begin{matrix} 
    \frac{t_i - t_0}{2 P f} & \mbox{when }0 \leq \frac{t_i - t_0}{P} < f\\ 
    \frac{t_i - t_0 - P}{2 P \left(1-f\right)}+1 & \mbox{when }f \leq \frac{t_i - t_0}{P} < 1
    \end{matrix}\right. 
\end{equation}
where $t_0$ is the Julian Date of the first observation, $P$ is the period, and $f$ is the asymmetry factor that has values between 0 and 1 and describes a steeper (f\,$<$\,0.5) or shallower (f\,$>$\,0.5) rising branch in comparison with the descending branch. Some NIR light curves have less than 10 points. In these cases, we used a simpler model with a symmetric cosine function (f\,=\,0.5) instead. We did not attempt to fit light curves with less than five observational points.

We probed periods $P$ between 100 and 2500 d in steps of 1 d, and in the case of asymmetric models we probed asymmetry factors $f$ between 0.10 and 0.90 in steps of 0.01. For each pair of values ($P,f$), we obtained best-fitting amplitudes $A$ and mean magnitudes $\overline{m}$ and their errors, by minimizing the sum of the squared residuals $\chi^2\equiv\chi^2\left(P,f\right)$. The residuals are the differences between observations and model, with the observations weighted with the inverse of the square of the photometric error. The minimum $\chi^2_{\rm min}$ of the $\chi^2$ values was used to select the most likely period and asymmetry factor (if applicable), together with the associated amplitude and mean magnitude for each of the four bands (optical, $J$, $H$, and $K$), when available. The error of the period $\sigma_P$ was obtained from $\sigma_P$\,$\simeq$\,$P^2$/$2T$, where $T$ is the time baseline of the observations (see \citealt{Stellingwerf78,Nakashima00}).

\begin{figure}
\includegraphics[width=\linewidth]{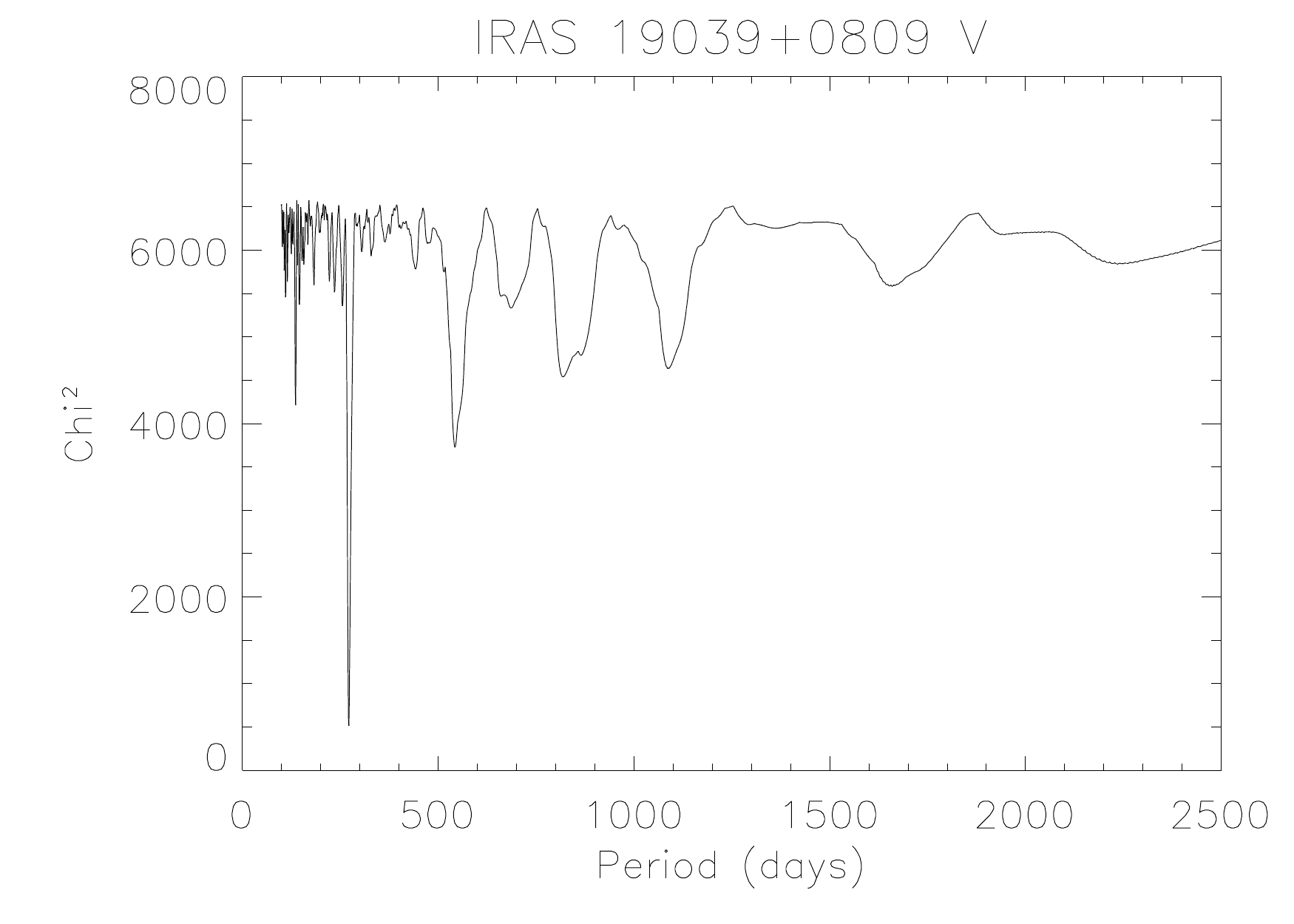}
\includegraphics[width=\linewidth]{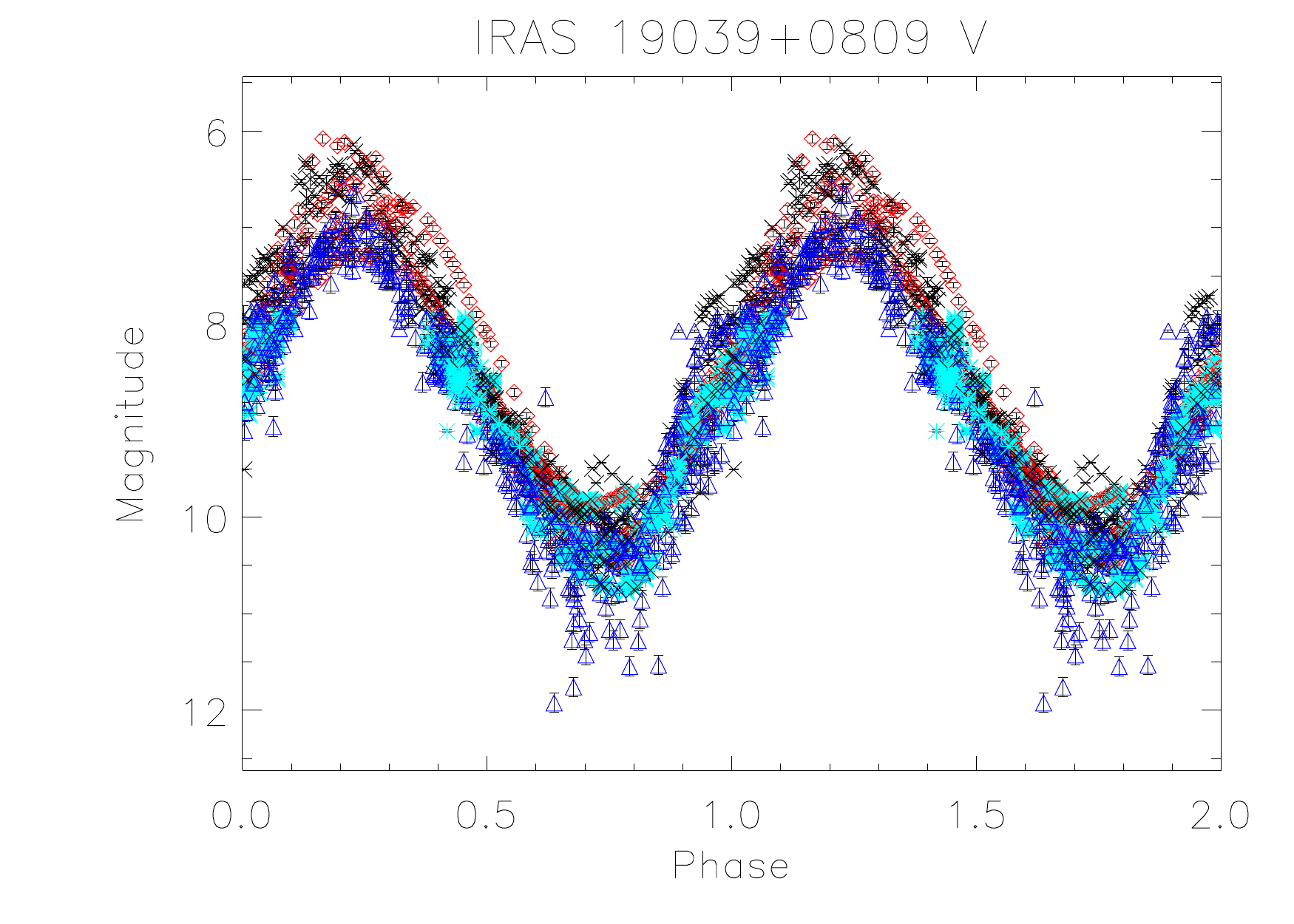}
  \caption{The periodogram (upper panel) and the folded light curve (lower panel) resulting from the fit of the optical light curve of R Aql (IRAS\,19039+0809). The period of 272 d is easily identified in the periodogram as the one of the (by far) smallest $\chi^{2}$ value. The folded light curve using this period is almost symmetric, with $f$\,=\,0.48. The dispersion of the observational data along the light curve reflects the typical cycle-to-cycle variation of Mira stars, clearly visible in Fig.\,\ref{F.19039LC}. Symbols in the lower panel are as in Fig.\,\ref{F.19039LC}.
  }
  \label{F.19039Fit}
\end{figure}

The results of the fits were visually inspected. Figure\,\ref{F.19039Fit} shows $\chi^2$ as a function of period (hereafter, periodogram) and the folded light curve resulting from the fit of the optical light curve of R Aql (IRAS\,19039+0809). In the periodogram it is possible to identify not only the best period of 272 d, but also several aliases. 

Mainly due to the scarcity of photometric points to populate the light curves, especially in the NIR, sometimes alias periods can present similar, or even lower, values of $\chi^{2}$ than the real period. The severeness of aliasing depends on the cadence of the observations, the regularity of the light curve, and how the observations are spaced along the variability cycle and is therefore difficult to quantify. Its relevance can be evaluated from the periodograms for each fit presented in Appendix\,\ref{Ap.Fits}. Thus, since it was expected that some of the automatically selected periods were not the correct values but aliases, we proceeded with three quality checks.

\subsection{Homogenizing solutions from different bands} 

First, in the case that an object had light curves in different bands, we used all of them to determine a consistent period. In most of the cases, individual periods obtained at different bands were compatible within the errors. For the cases with inconsistent individual periods among the different bands, we proceeded with a second visual inspection of the periodograms to identify another period compatible with the fits in all bands. Then, we repeated the fits with a search window of $\pm$50 d around the selected new period, and confirmed the solutions by visual inspection of the new folded light curves. In cases where a common period could not be determined, we kept the individual period of the light curve judged as the most reliable one, and we rejected the other solutions.

\subsection{Evaluation of amplitude ranges}

Our second quality check was based on the analysis of the amplitudes of the fit solutions. Occasionally, the fits resulted in implausible small or large amplitudes. In these cases, other individual periods were tried looking for reasonable amplitudes, but always guaranteeing compatible periods in all analysed bands. In the case that no solution with reasonable amplitude was identified, we kept the initial solution if the fit was reliable despite of the anomalous amplitude. Otherwise, all solutions were rejected assuming that no good solution was possible with the current data.

\subsection{Combining periods from different bands}

Once the individual periods obtained from the different bands were compatible, we averaged them for each source, weighting each individual period with its error divided by the number of observations in the corresponding light curve. We conservatively defined as error of the combined period, the larger between the standard deviation and the minimum error of the individual periods. In total, we could determine combined periods for N\,=\,345 Arecibo sources.

\subsection{Comparison with periods from the literature}
\label{Sect.comparation}

The third and last quality check was done by comparing our results with light curves and period estimations publicly available for part of the Arecibo sample. We did an extensive search among the optical public surveys ASAS-SN, ATLAS \citep{Heinze18}, Catalina \citep{Drake14}, \emph{Gaia} DR2, and KELT \citep{Arnold20}, in the General Catalogue of Variable Stars (GCVS; \citealt{Samus17}), and in the literature \citep{Herman85a,Carter92,LeBertre93,Whitelock94,Tang08,Usatov08,Engels15a,Vogt16,Urago20}, obtaining periods for 193 ($\sim50$\%) objects in our sample. 

When more than one period was available from the literature, we verified the compatibility of the values, and in case of clear differences we inspected the light curves, when available, and selected the one that we considered as the most reliable. In five cases the literature periods belong to sources without period determinations in this work.

\begin{figure}
    \includegraphics[width=\linewidth]{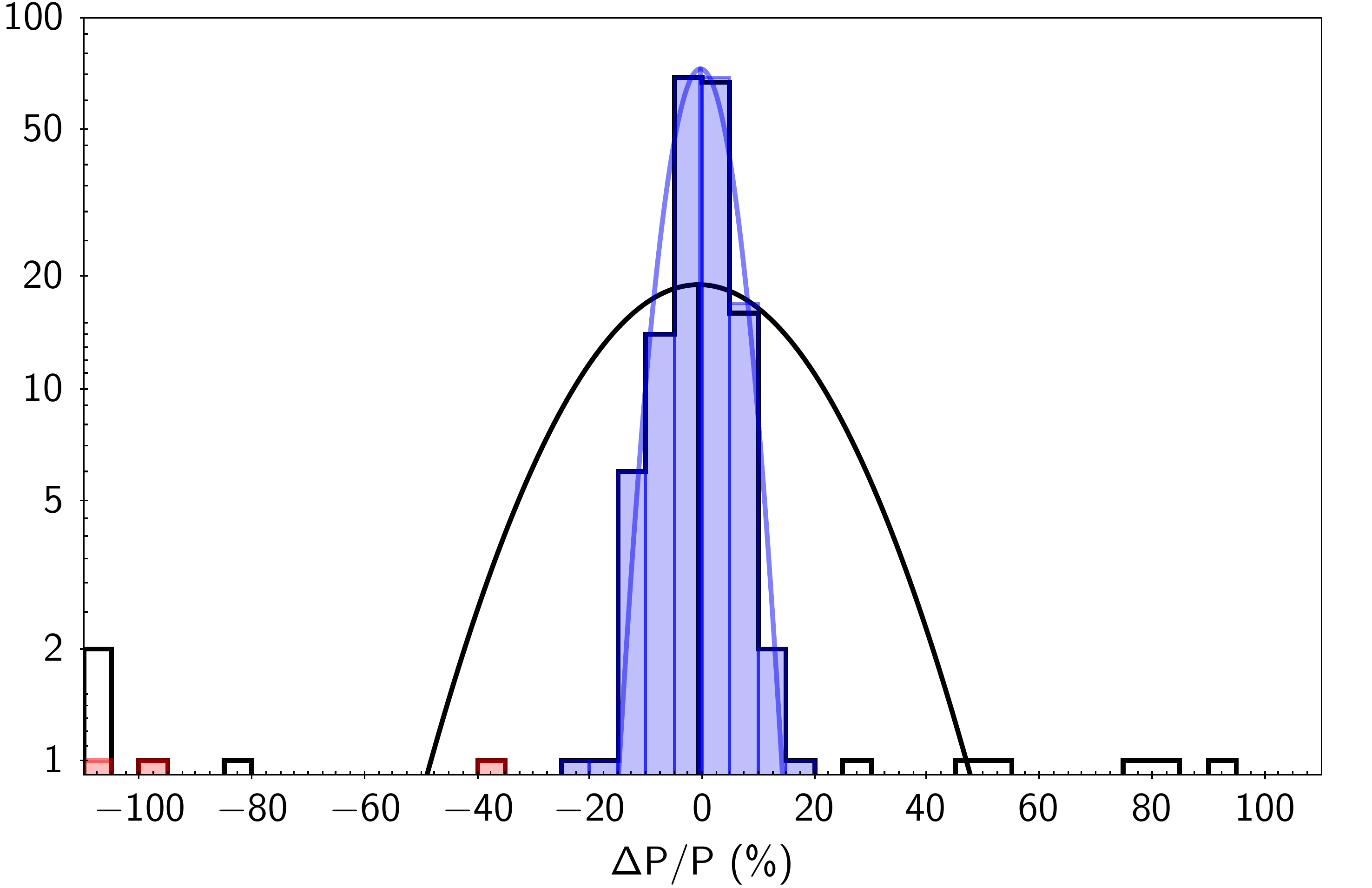}
  \caption{Distribution of the differences between our combined periods and those found in the literature, before (black) and after (blue) corrections (see the text). Solid lines are Gaussian fits to the distributions. The three cases, where the combined periods were revised in favor of the literature periods are shown in red. We use a logarithm scale in the $Y$-axis to better show the outliers.}
  \label{F.Diff-P}
\end{figure}

Figure\,\ref{F.Diff-P} shows the distribution of the differences $\Delta P$ between the most reliable periods collected from the literature and our combined periods for the 188 sources in common, expressed as percent of the combined periods (black histogram). The solid black line is a Gaussian fit to the $\Delta P$/P distribution, with a mean ($\Delta P$/$P$) of $\sim 0$ and a standard deviation of $\sim 20$\%. Thus, the periods of 177 (94\%) sources are compatible with those of the literature, which validates our method to determine the variability properties of our sample. The reasons for the periods deviating by more than $\sim 20$\% (11 sources) are discussed in the following:

\begin{itemize}
    \item Five of the discrepant literature periods were from the ASAS-SN survey. Three of them (IRAS\,18592+1455, IRAS\,19256+0254, and IRAS\,21305+2118) correspond to very obscured objects in the optical, so very probably the ASAS-SN data belong to a contaminating nearby field star. IRAS\,02404+2150 has a period of 184 d in ASAS-SN, which coincides with an alias of the more reliable solution of $\sim 368$ d from the NIR-DB photometry (see Appendix\,\ref{Ap.Fits}). In the case of IRAS\,19343+0912, ASAS-SN provides a period of $\sim 250$ d, which is clearly wrong based on the folded light curve published on the ASAS-SN web page. For this source we tried our own fit with combined optical data from ASAS-SN and ASAS and found no periodicity, although there is clear variability in the optical data. Summarizing, in all five cases we consider our combined period as the correct one.
\item Another three discrepant periods were from the ATLAS survey. For IRAS\,19344+0016, ATLAS provided a period of $\sim$\,800 d, while we got $\sim$\,450 d. Our own fit to the ATLAS data resulted in $P$\,$\sim$\,430 d, consistent with our combined period. In the case of IRAS\,19464+3514, ATLAS provides a period of $\sim 1100$ d, but we were not able to verify it with our own fit procedure. This is a very red source with no optical counterpart in the DSS; therefore, we consider the ATLAS period as unreliable. For the third source, IRAS\,20115+0844, the ATLAS period is 429 d, about twice as long as the combined period from the NIR-DB photometry. We verified the ATLAS period by performing our own fit to the ATLAS data. Since in each of the periodograms of our NIR-DB light curves a clear local minimum is present close to the ATLAS period (see Appendix\,\ref{Ap.Fits}), we forced our fits to a solution around this value and determined a new combined period of $P=429\pm42$ d.
    \item Two more sources with discrepant literature periods were from an (ongoing) OH maser monitoring program\footnote{\url{https://www.hs.uni-hamburg.de/nrt-monitoring}} \citep{Engels15a}. IRAS\,19283+1944 (OH~55.0+0.7) has an OH radio period of $\sim 1270$ d. The periodogram of this source shows a second minimum around this value (see Appendix\,\ref{Ap.Fits}). Since we considered the radio period as more reliable because the OH light curve is better sampled, we obtained a new combined period of $P=1275\pm244$ d based on the NIR-DB data. In the case of IRAS\,19576+2814 (OH~65.7--0.8), we considered the period of $\sim 1500$ d from the OH monitoring as tentative. In addition, there is no minimum in the periodogram of this source around this value (see Appendix\,\ref{Ap.Fits}), so we decided to keep our result ($P \sim 2000\pm600$ d). 
\item Finally, IRAS\,19352+1914 has a period in the literature of $\sim 860$ d \citep{Urago20}, while our fit provided a value of $\sim 620$ d. The periodogram of this source also shows a local minimum around 900 d that corresponds with the literature period. Since the light curve from the literature is better sample than ours, we adopted $P=891\pm129$ d for the combined period based on our NIR-DB.
\end{itemize}


Thus, of the 11 discrepant literature periods, 3 led to corrections of our initially determined combined periods. They are marked in red in Fig.\,\ref{F.Diff-P}. The Gaussian fit to the distribution after corrections, shown in blue in Fig.\,\ref{F.Diff-P}, has a standard deviation of $\sim 5$\%. Removing the eight erroneous literature periods from the comparison, we find that 177 out of 180 combined periods were confirmed ($\sim 98$\%). Among the combined periods derived from poorly sampled NIR light curves ($<10$ observations) aliasing problems may still have escaped our verification method. However, there are only four sources with such light curves for which no optical variability data or periods from the literature are available for comparison. Based on these considerations, we conclude that among in total 345 determined combined periods, only a handful ($<3$\%) are expected to be erroneous.

Among the five periods found in the literature, which belong to the remaining 40 objects without determination of combined periods, we added the periods of IRAS\,01037+1219 \citep{Drake14}, and IRAS\,18498$-$0017 and 19067+0811 \citep{Engels15a} to our sample, so that in total final periods are available for N\,=\,348 Arecibo sources.

We present in Table\,\ref{t:fitsummary} the value of the different parameters that define each light curve, and the final periods accompanied by a field that refers to the data used to obtain the combined period: $O$ for optical; $J$, $H$, and $K$ for each NIR band; and `Lit' for literature origin. The full table is available in an electronic format (see Data Availability).

\begin{table*}
\begin{center}
\renewcommand{\tabcolsep}{2.0pt}
\centering
\caption{Results of the fits to the NIR and optical light curves. The periods `P' are weighted averages of the periods in the individual bands listed in `Bands' or are from the literature (`Lit.'). `Type' is the light-curve model used (A\,=\,Asymmetric, S\,=\,Symmetric), $f$ is the asymmetry factor, and t$_0$ is the date of the first photometric data. The whole table is available electronically.} 

\label{t:fitsummary}
\scalebox{0.8}{
\rotatebox{90}{
\begin{tabular}{llcc|ccccccc|ccccccc|ccccccc|ccccccc}
\hline \hline\noalign{\smallskip}
& & &\,\,\,& \multicolumn{6}{c}{Optical} &\,\,\,&\multicolumn{6}{c}{J} &\,\,\,& \multicolumn{6}{c}{H} &\,\,\,& \multicolumn{6}{c}{K}\\
\cline{5-10} \cline{12-17} \cline{19-24} \cline{26-31} 
\noalign{\smallskip}
IRAS & Bands & P && Type & P$_{O}$ & A$_{O}$ & $\overline{m_O}$ & $f_{O}$ & t$_{0_{O}}$ && Type & P$_{J}$ & A$_{J}$ & $\overline{m_J}$ & $f_{J}$ & t$_{0_{J}}$ && Type & P$_{H}$ & A$_{H}$ & $\overline{m_H}$ & $f_{H}$ & t$_{0_{H}}$ && Type & P$_{K}$ & A$_{K}$ & $\overline{m_K}$ & $f_{K}$ & t$_{0_{K}}$  \\
 & & (days) && & (days) & (mag) & (mag) & & && & (days) & (mag) & (mag) & & && Type & (days) & (mag) & (mag) & & && Type & (days) & (mag) & (mag) & &  \\
\hline\noalign{\smallskip}
01037+1219 & Lit & 650  \\
01085+3022 & JHK  & 526$\pm$63 &&   &             &      &      &      &         && S & 559$\pm$85 & 1.8 & 5.0 & --- & 2451367 && S & 568$\pm$88 & 1.7 & 3.6 & --- & 2451367 && S & 483$\pm$63 & 1.6 & 2.2 & --- & 2451367\\
02404+2150 & O &    368$\pm$44 && A & 368$\pm$44  & 0.45 & 13.8 & 0.23 & 2456630 \\
02420+1206 & O &    354$\pm$10 && A & 354$\pm$10  & 5.37 & 13.0 & 0.24 & 2452454 \\
02547+1106 & OJHK & 459$\pm$63 && A & 450$\pm$187 & 2.71 & 13.6 & 0.23 & 2456905 && S & 461$\pm$63 & 1.6 & 5.8 & --- & 2451529 && S & 461$\pm$63 & 1.4 & 4.6 & --- & 2451529 && S & 467$\pm$65 & 1.3 & 3.8 & --- & 2451529\\
\noalign{\smallskip}\hline
\end{tabular}
}
}
\end{center}
\end{table*}


\section{Discussion}
\label{sec:discussion}

The Arecibo sample of OH/IR stars contains stars with a wide range of IRAS colours (cf. Fig. \ref{fig:IRAScc}) including optically bright Mira variables at the blue end of the [12]$-$[25] colour range and extreme OH/IR stars, visually obscured by the dust in the CSE, at the opposite end (Paper I). The latter are a mixed group of long-period large-amplitude variable (LPLAV) stars at the end of the AGB and non-variable post-AGB stars still emitting OH maser. Thus, in addition to the predominant LPLAV, a smaller number of semiregular or irregular variable AGB stars, post-AGB stars, and Red Supergiants (RSGs) are expected to be contained in the Arecibo sample. Among the very red sources also contamination with YSOs is possible \citep{Garcia-Lario97}.

\subsection{Arecibo source classification}
\label{sec:classification}

On an individual basis, a classification is now possible with the help of their variability properties together with an extensive literature search using ADS\footnote{SAO/NASA Astrophysics Data System, \\ 
\url{https://ui.adsabs.harvard.edu/}} 
and SIMBAD\footnote{SIMBAD Astronomical Database - CDS (Strasbourg), \\ 
\url{http://simbad.u-strasbg.fr}}. Thus, the Arecibo sources were broadly distributed into four groups. The largest group are the LPLAVs. Its 345 sources have large amplitudes typical for Mira variables and variable OH/IR stars, qualifying them as AGB stars. A second group is made up of 16 post-AGB star (candidates), from which seven were already identified in Paper\,I\footnote{Note that IRAS\,19178+1206 was misidentified in Paper\,I with a field star and was erroneously classified as non-variable OH/IR star. The object was not detected during our observations and is still considered as `unclassified' (cf. Table\,\ref{t:no-phot}).}. Another group consists of 11 sources for which there is no variability information available, and so remain unclassified. The fourth group is formed by the rest of sources with a variety of types. The group membership of each source is listed in the column `Group' of Table\,\ref{t:sample}. The 29 sources classified different than LPLAV (Groups 2 and 4) are listed in Table\,\ref{T:no-ohir}.

\subsection{Variability properties of LPLAVs}

\begin{figure}
    \includegraphics[width=\linewidth]{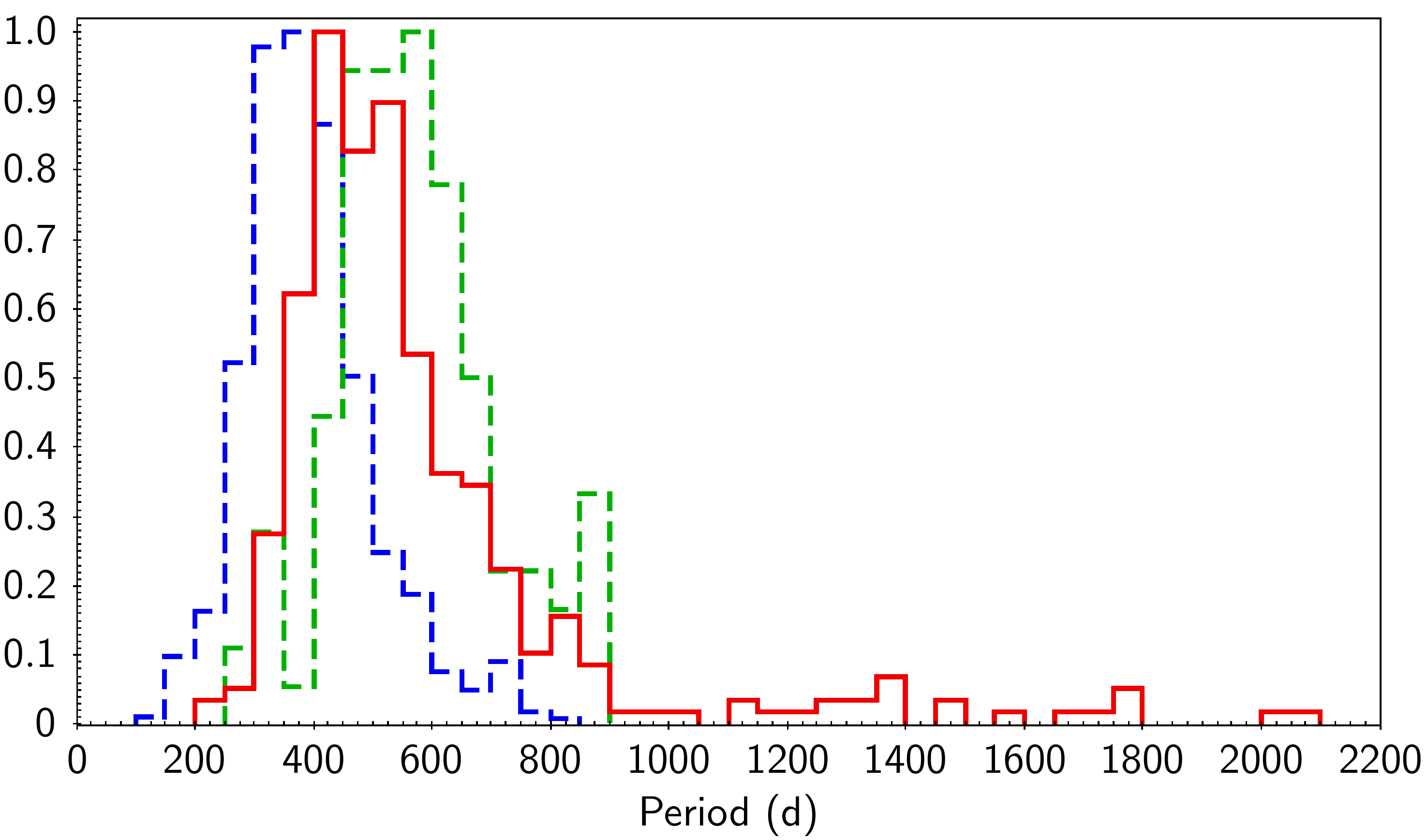}
  \caption{Distribution of the periods for the Arecibo sample (solid red), the O-rich AGB stars compiled by \citet{Suh17} (dashed blue), and the sample of LPVs from \citet{Urago20} (dashed green). All distributions are normalized to their maximum value.}
  \label{F.Periods}
\end{figure}

Our light-curve analysis confirmed that the large majority of the Arecibo sample ($\sim$90\%) are LPLAVs. They are O-rich Mira variables and extreme OH/IR stars. Figure\,\ref{F.Periods} shows the period distribution of the Arecibo sample. Most of the sources present variability in the range from $\sim 300$ to 800 d. The minimum period in the sample is 223 d and longer periods are distributed along a long tail up to $\sim 2100$ d.

In addition to the Arecibo sample, we show in Fig.\,\ref{F.Periods} the distribution of periods of almost 1500 O-rich AGBs compiled by \cite{Suh17}, and the sample of 108 long-period Mira variables in the Galactic disc presented by \cite{Urago20}. All distributions were normalized to their maximum value. 

The Arecibo sample contains longer periods than Suh's sample, which only contains periods $P<900$ d. Since \cite{Suh17} compiled the sample from the GCVS \citep{Samus17} and the AAVSO, it is biased towards the optically bright Mira variables, while the Arecibo sample also contains many optically obscured sources. On the other hand, the decrease of the number of sources with periods shorter than $\sim 400$ d in the Arecibo sample is caused by the IRAS colour-based selection of the sample, which excluded part of the Mira population with relatively blue colours. As on average periods decrease with bluer IRAS colours (see Section \ref{sec:pc-correlations}), Mira variables with shorter periods are underrepresented in the Arecibo sample.

The period distribution of the sample of \cite{Urago20} peaks between 450 and 650 d. Although it does not contain periods $P>900$ d either, the distribution peaks at longer periods than the period distribution of the Arecibo sample ($\sim 400-600$ d; see Fig. \ref{F.Periods}). Urago et al. selected their sample from boxes IIIa and IIIb in the IRAS 2CD (Fig. \ref{fig:IRAScc}), overlapping to a great extent with the colours of the Arecibo sample. However, from their list of sources published we have only 37 sources in common, all with compatible periods except one (see Section \ref{Sect.comparation}). The differences in the period distributions are due to selection effects. First, the lack of periods $P>900$ d is due to the specific sources selected for the publication (Urago, private communication). Second, the IRAS colour distributions differ between both samples. The sample of Urago et al. is split in almost equal parts into blue and red sources (we set the division line at [12]$-$[25]\,=\,0.1 mag), while in the Arecibo sample the blue sources pre-dominate (64\%). The maximum of the period distribution of the sample of Urago et al. is therefore shifted to longer periods, because of the correlation between periods and infrared colours (see Section \ref{sec:pc-correlations}).

\begin{figure}
    \includegraphics[width=\linewidth]{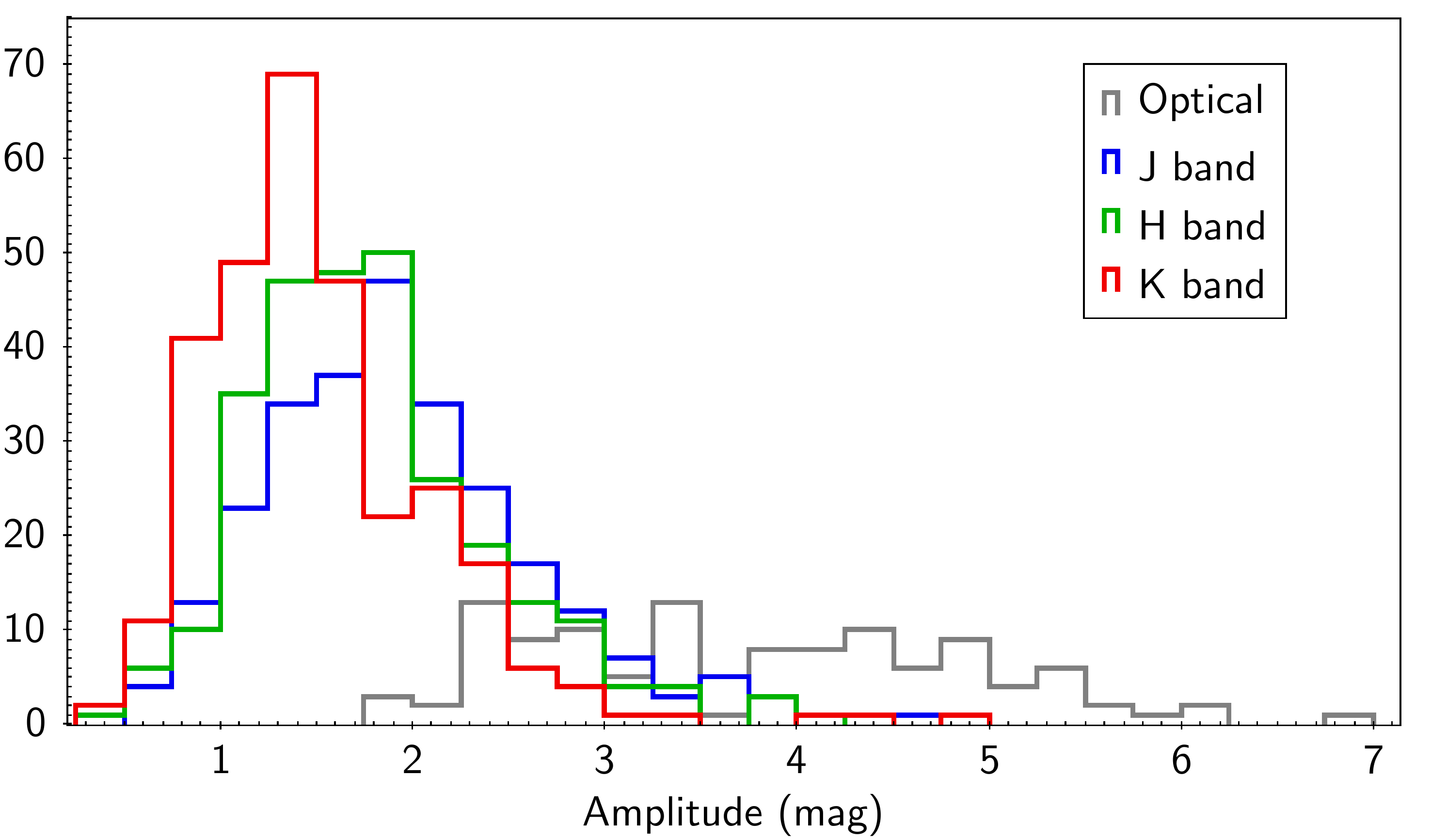}
  \caption{Distributions of peak-to-peak amplitudes for the Arecibo sample in the optical band and the three NIR bands as obtained from the light-curve fit (see Section \ref{sec:LC-fit}).}
  \label{F.Amplitudes}
\end{figure}

The Arecibo sample presents a large range of brightness variations. Figure\,\ref{F.Amplitudes} shows the distribution of the peak-to-peak amplitudes obtained from the light-curve fits. Typically, amplitudes are between 1 and 3 mag in the NIR, and between 2 and 6 mag in the optical. Amplitudes decrease from shorter to longer wavelengths.

\begin{figure}
    \includegraphics[width=\linewidth]{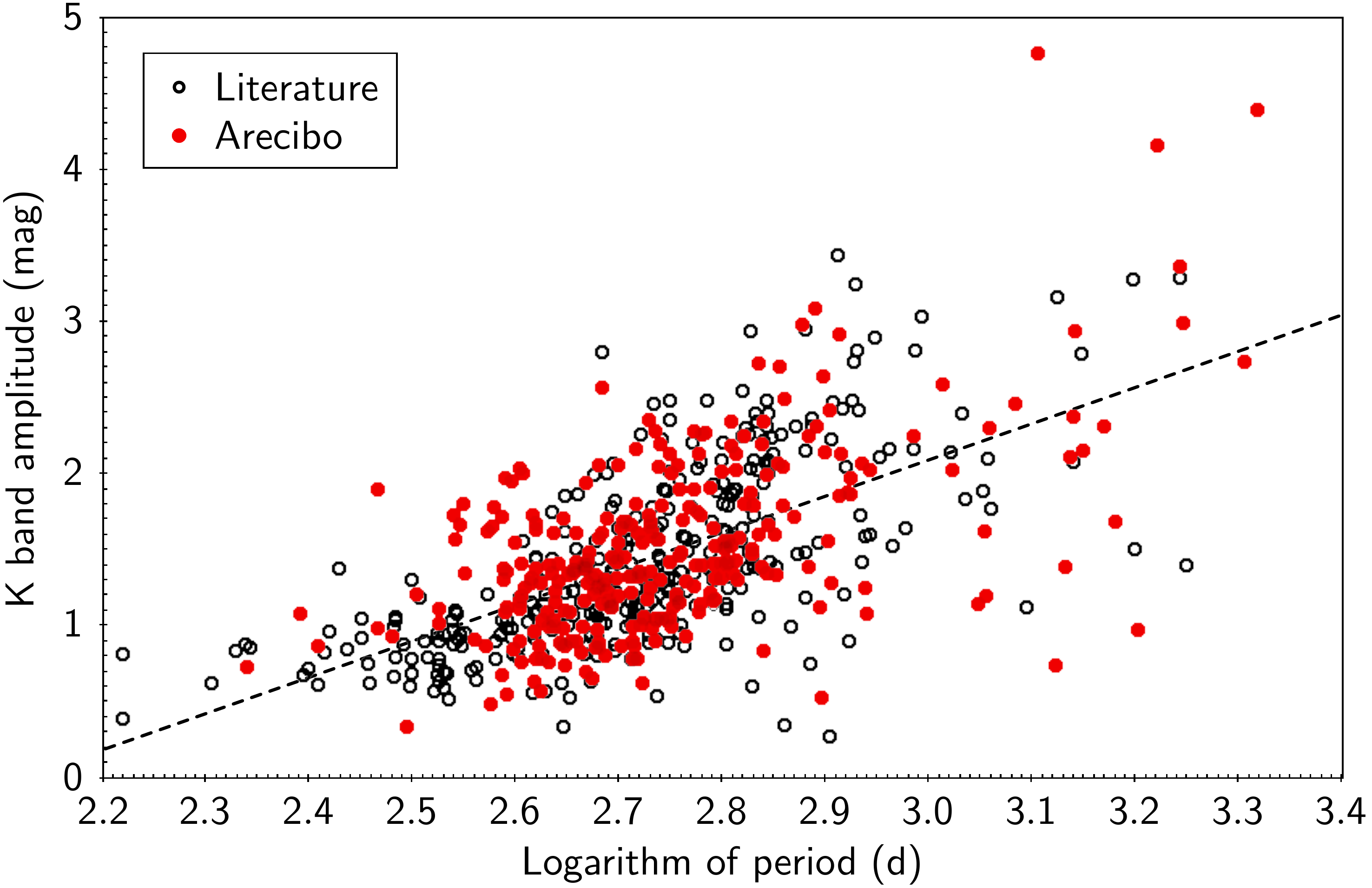}
  \caption{Peak-to-peak amplitudes in the $K$ band as a function of the period from our NIR-DB photometry (red filled circles) and from the literature (black open circles). Dashed line is a linear fit to all data (see the text).}
  \label{F.P_vs_Amp}
\end{figure}

Figure\,\ref{F.P_vs_Amp} shows the relation between peak-to-peak amplitudes and periods obtained from the $K$-band light-curve fit. A compilation of 336 periods and $K$-band amplitudes from the literature is also included \citep{Engels83,Jones90,LeBertre93,Whitelock94,Wood98b,Olivier01,Jimenez-Esteban06a,Tang08,Urago20}.

The distribution of amplitudes against periods in the Arecibo sample matches very well the distribution obtained from the literature, which supports the validity of our light-curve fitting results. As expected (\citealt{Jimenez-Esteban06a,Urago20}), the amplitude shows a correlation with the period in the sense that longer periods are connected to larger amplitudes. We fitted a line using all available data, which follows the equation
\begin{equation}
    A_K\,=\,2.38 log\left(P\right)-5.06
\end{equation}
where A$_K$ is the peak-to-peak amplitude in the K band in magnitudes and P is the period in days. 

\begin{figure}
    \includegraphics[width=\linewidth]{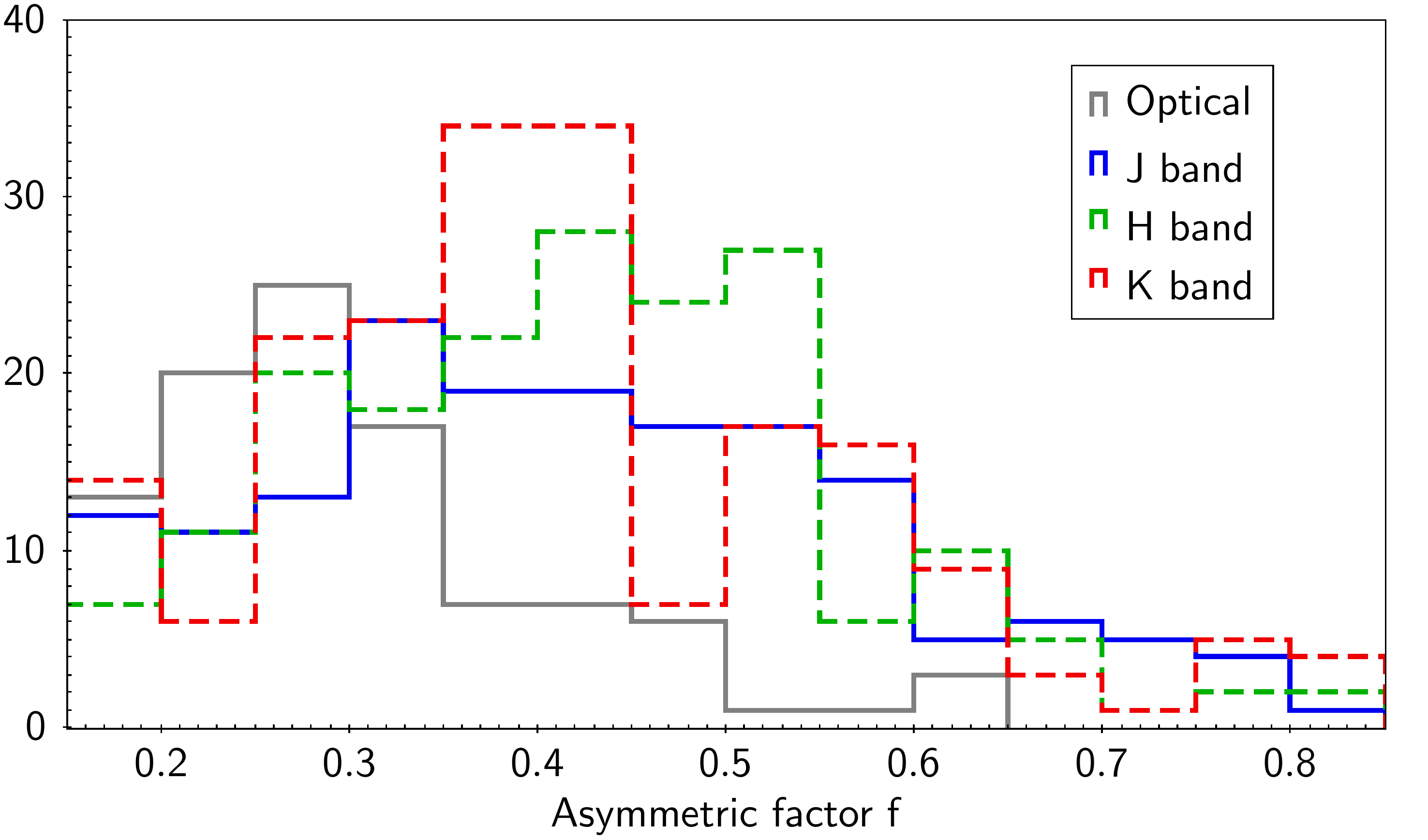}
  \caption{Distributions of the asymmetry factor $f$ for the Arecibo sample in the optical band and in the three NIR bands as obtained from the light-curve analysis.}
  \label{F.AsyFact}
\end{figure}

Figure\,\ref{F.AsyFact} shows the distribution of the asymmetry factor $f$ for the four bands used in the light-curve analysis. Most of the light curves have $f$\,$\la$\,0.5, which correspond to a steeper rising branch in comparison with the descending branch. This is a common property of Mira light curves (\citealt{Vardya88} and references therein; \citealt{Lebzelter11}). The asymmetry factor was not correlated with period or with amplitude.

\subsubsection{Variation along the luminosity cycle \label{var-cycle}}

We used the light curve of IRAS\,19454+2536 to analyse the variation of the colour along the variability cycle. This source has one of the best sampled light curves with observations at the maximum and minimum of its cycle (see Appendix\,\ref{Ap.Fits}). In Fig.\,\ref{F.colour-cycle}, we show its $J$-band light curve (upper panel), and the variations of the NIR $J-K$ colour during the same period of time. There is a clear change in colour along the pulsation cycle, with a variation of $\sim 1$ mag. The bluer colours, so the higher temperatures, correspond with the epoch of maximum luminosity and the redder colours, so the lower temperatures, with the epoch of minimum luminosity. The change of the photometric properties along the variability cycle provides important constraints to derive the physical properties of the emitting material \citep[e.g.][]{Bladh13}.

\begin{figure}
\includegraphics[width=\linewidth]{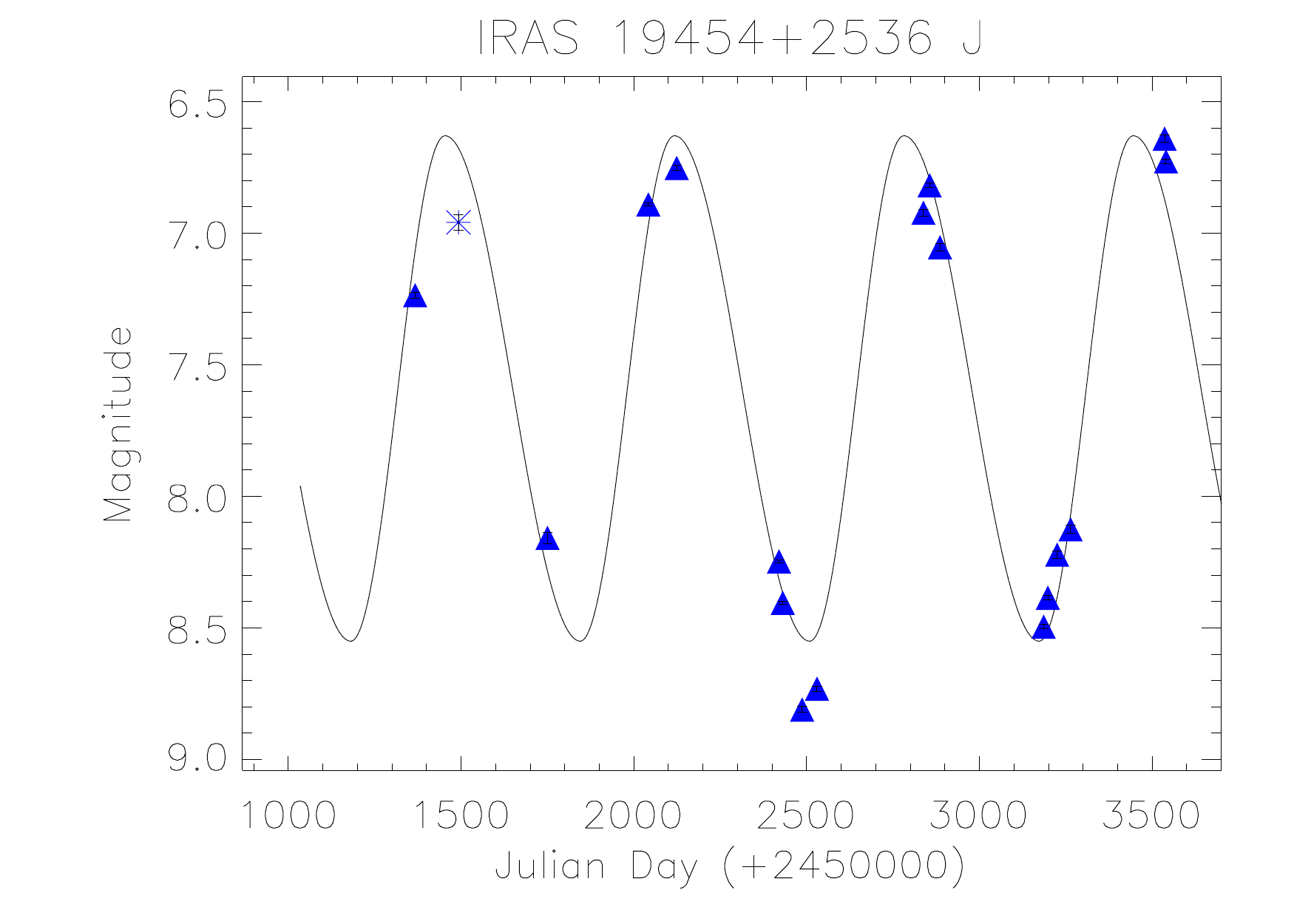}
\includegraphics[width=\linewidth]{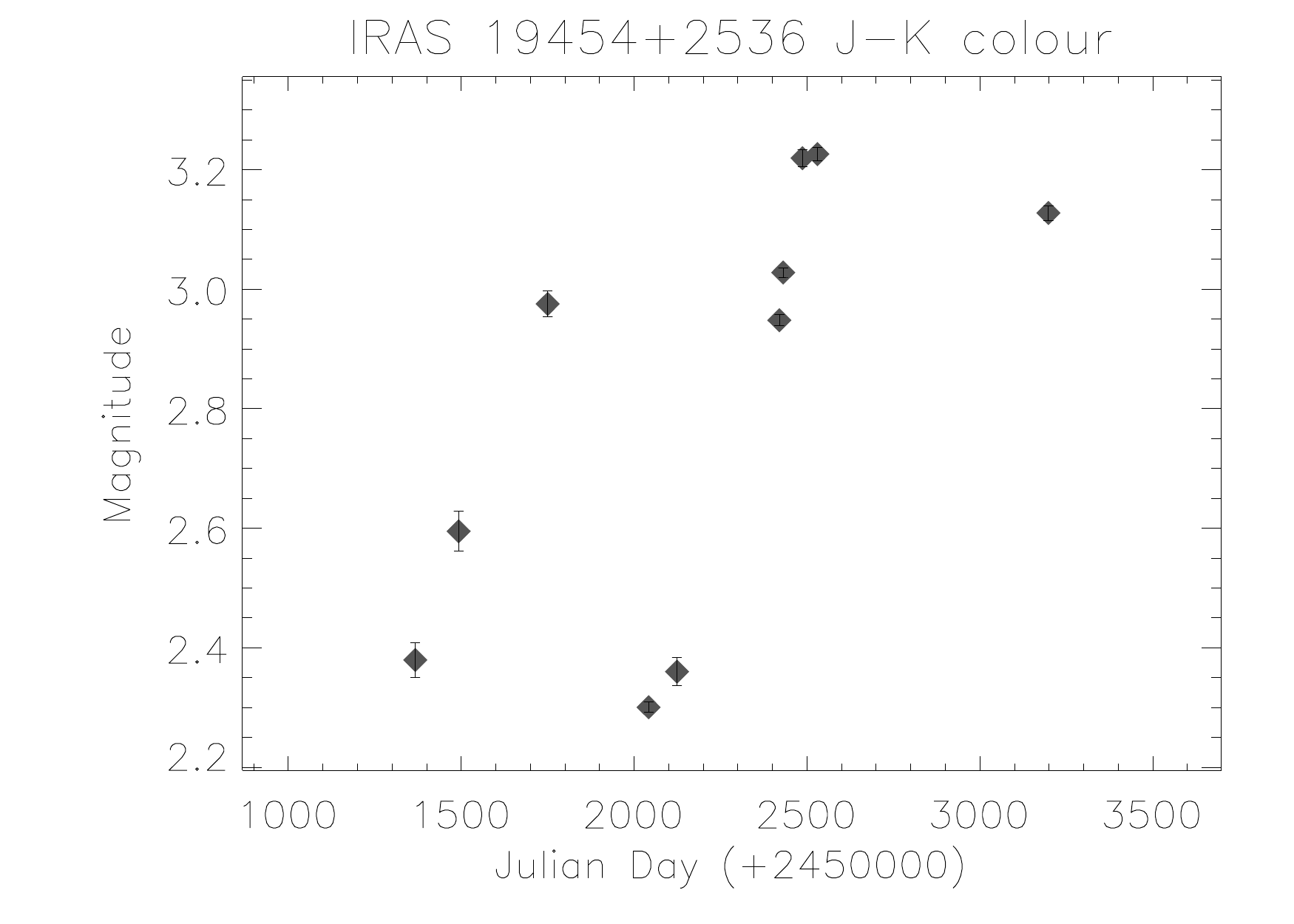}
  \caption{$J$-band light curve of IRAS\,19454+2536 with a variability period of 664 d (upper panel) and the $J-K$ NIR colours at the same epochs (lower panel). Photometric errors are shown with bars. The lower number of $J-K$ colours is due to the lack of $K$-band photometry in several epochs.}
  \label{F.colour-cycle}
\end{figure}

\subsubsection{Period--colour correlations}
\label{sec:pc-correlations}

\begin{figure}
    \includegraphics[width=\linewidth]{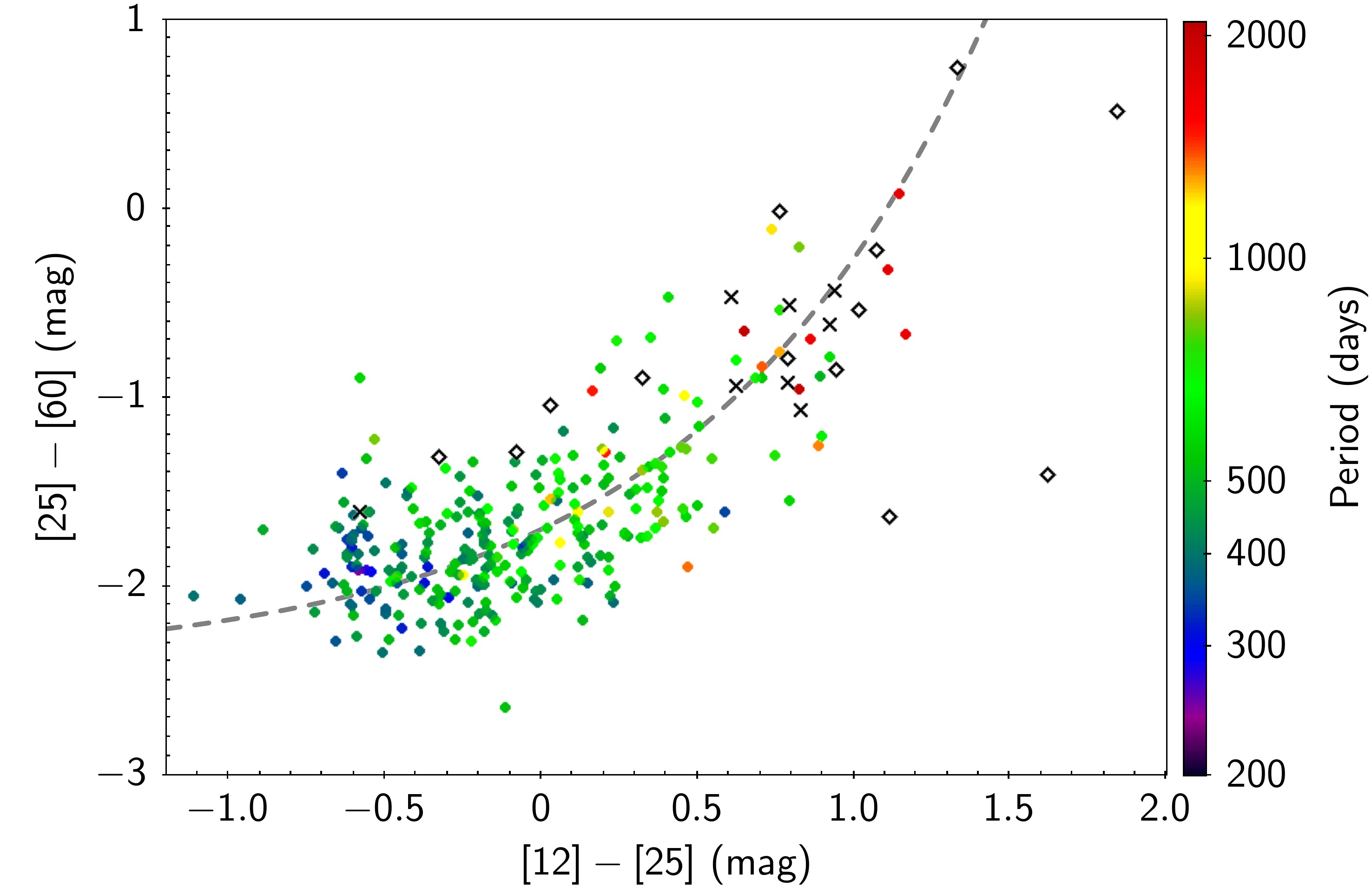}
    \includegraphics[width=\linewidth]{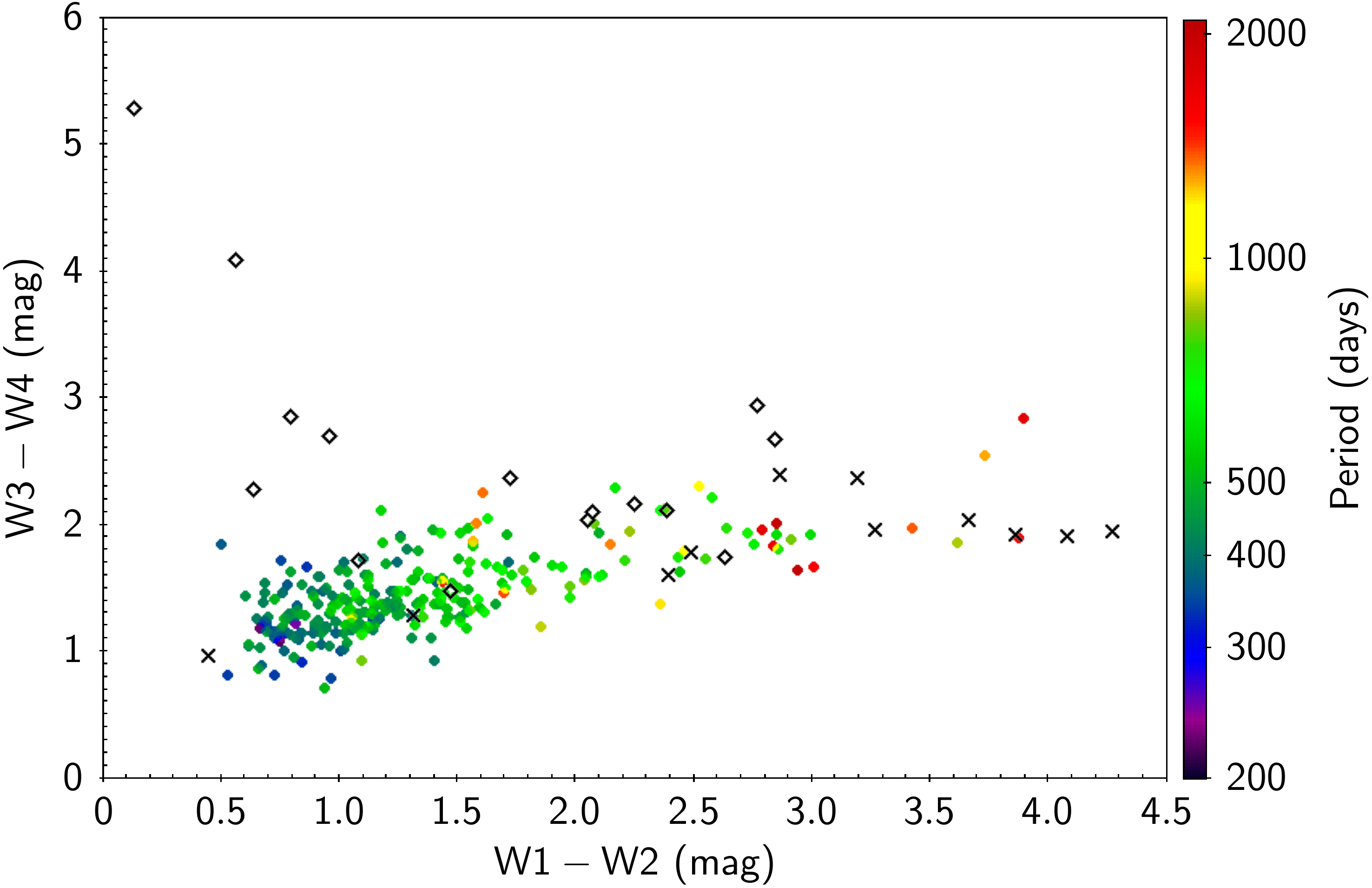}
    \includegraphics[width=\linewidth]{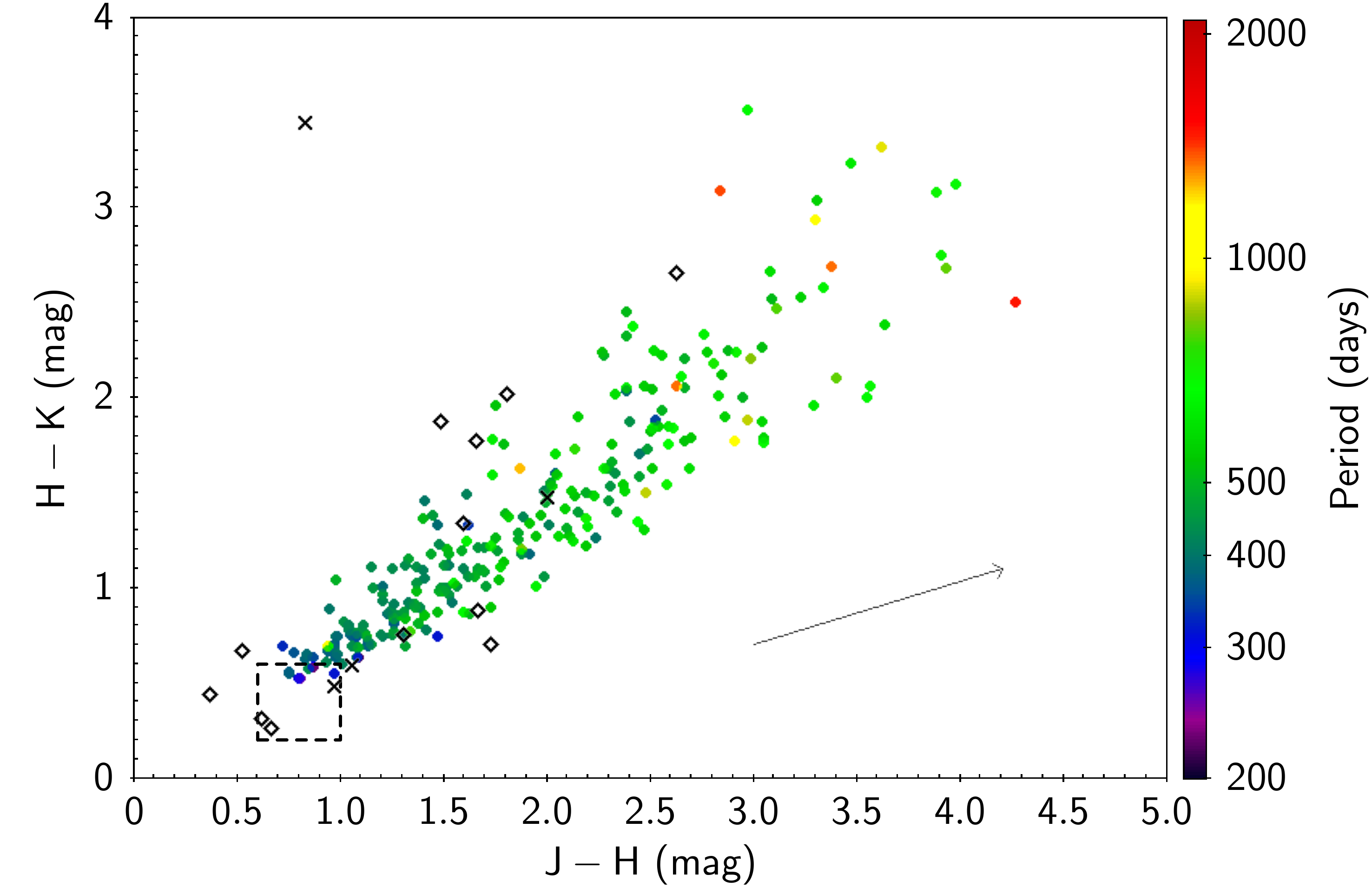}
    \caption{Mid/far-IR (top), near/mid-IR (middle), and near-IR (bottom) 2CDs of the Arecibo sample colour coded by period. LPLAVs are shown with full circles, post-AGBs with open diamonds, and unclassified sources with crosses. We do not show sources with upper limits in any band. The dashed gray line in the top panel is the `oxygen-rich AGB sequence' (see Section \ref{sample}). The arrow in the bottom panel is the A$_V$\,=\,10 mag reddening vector \citep{Cardelli89}, and the dashed line box is the region occupied by optically selected Mira variable stars.}
  \label{F.IRcc}
\end{figure}

The correlation between the variability period and the infrared colours of the Arecibo sample is studied in this section. We show in Fig.\,\ref{F.IRcc} three infrared 2CDs: The mid/far-IR IRAS 2CD is at the top panel, the near/mid-IR AllWISE 2CD is at the middle panel, and the NIR 2CD is at the bottom panel. Arecibo LPLAVs are shown with full circles, which have been colour coded based on the period value. Additionally, we show in Fig.\,\ref{F.IRcc} the IR colours of the Arecibo sources classified as small-amplitude non-periodic post-AGB variables (Group\,2; open diamonds), and those unclassified (Group\,3; crosses). We only show sources with detections in the IRAS bands (Quality flag 2 and 3) and in the AllWISE bands (Quality flag A, B and C). For the NIR 2CD we used the mean magnitude obtained from the light-curve fits for the LPLAVs (Table\,\ref{t:fitsummary}), and the mean magnitude obtained from the observational data for the rest (Table\,\ref{t:photosummary}). 

The Arecibo sample presents NIR colours much redder than typical Mira variables selected in the optical. The dashed line box shows in Fig. \ref{F.IRcc} (bottom panel) the region occupied by such stars in the NIR 2CD \citep{Feast82}, which corresponds to the bottom end of the Arecibo NIR colour distribution. In addition, the Arecibo LPLAVs with the longest periods ($\ga$\,1500 d) are not present in the NIR 2CD because they were not detected either in the $J$ or $H$ band during the NIR-MP. Some of them have the UKIDSS data, and their colours are among the reddest, with $H-K$ ranging between 3 and 6.5 mag.

In all three 2CDs, periods generally increase from bluer to redder colours. Such period--colour relations are well known for pulsating AGB stars in the Galaxy (\citealt{Engels83}; \citealt{Whitelock94}). A recent update using WISE colours including several thousand AGB stars in the Galaxy, as well as in the Large Magellanic Cloud, has been presented by \cite{Suh20}. The relation is also found by the \cite{GaiaCollaboration19-Eyer} in LPVs within 1 kpc distance from the Sun using \emph{Gaia} colours.

\subsection{Post-AGB stars and candidates}

   \begin{table*}
      \caption[]{List of sources classified as post-AGB stars (Group 2; N\,=\,16) or having miscellaneous classifications (Group 4; N\,=\,13). Because of their OH non-detection, the post-AGB stars IRAS\,18551+0159 and 19200+1035 were assigned to Group 4.}
      \label{T:no-ohir}
\begin{center}
\begin{tabular}[t]{lllll}
\hline\hline\noalign{\smallskip}
IRAS & Other name & Group and classification & Reference\\
\hline\noalign{\smallskip}
02404+2150 & YY\,Ari                  & 4 - Semiregular variable                                       & \cite{Whitelock95} \\     
05506+2414 &                          & 4 - Young stellar object                                                        & \cite{Sahai08} \\         
06238+0904 & RAFGL\,940               & 4 - Carbon star; questionable OH detection                     & \cite{Chen10} \\          
06319+0415 & RAFGL\,961               & 4 - Young stellar object                                       & \cite{Cooper13} \\        
07331+0021 & AI\,CMi                  & 2 - Post-AGB star                                              & \cite{Arkhipova17} \\     
18095+2704 & HD\,335675               & 2 - Post-AGB star                                              & \cite{Hrivnak88} \\       
18123+0511 & V2053\,Oph               & 2 - Proto-planetary nebula                                     & \cite{Lagadec11} \\       
18455+0448 &                          & 2 - Post-AGB star; water fountain                              & \cite{Vlemmings14} \\     
18520+0533 &                          & 2 - Post-AGB star candidate                                    & Paper\,I \\               
18551+0159 & GPSR\,035.472$-$0.436     & 4 - False OH detection; planetary nebula candidate             &  \cite{Zhu13} \\          
18577+1047 &                          & 2 - Proto-planetary nebula candidate                           & \cite{Lewis89} \\         
18596+0315 & OH\,37.1$-$0.8            & 2 - Post-AGB  star; water fountain                             & \cite{Gomez17} \\         
19035+0801 & PN\,G041.7+00.4          & 2 - Planetary nebula                                           & \cite{Kohoutek01} \\      
19065+0832 & OH\,42.6+0.0             & 2 - Post-AGB star candidate                                    & Paper\,I \\               
19083+0851 &                          & 4 - RSG star                                        & \cite{Engels96} \\        
19177+1333 &                          & 2 - Post-AGB star candidate                                    & This work \\              
19189+1758 &                          & 4 - Optical small amplitude regular variability                & This work \\              
19200+1035 & PN\,K\,3$-$33             & 4 - False OH detection; planetary nebula                       & \cite{Acker92} \\         
19229+1708 &                          & 4 - RSG star                                        & \cite{Winfrey94} \\       
19254+1631 & OH\,51.8$-$0.1            & 2 - Post-AGB star; water fountain candidate                    & \cite{Vlemmings14} \\     
19319+2214 &                          & 2 - Post-AGB star candidate                                    & This work \\              
19343+2926 &  Min\,1$-$92              & 2 - Bipolar proto-planetary nebula                             & \cite{Bujarrabal98} \\    
19500+2239 &                          & 4 - Irregular variability                                      & This work \\              
19566+3423 &                          & 2 - Proto-planetary nebula                                     & \cite{SanchezContreras12} \\     
20015+3019 & V719 Cyg                 & 4 - RSG star                                        & \cite{Winfrey94} \\       
20127+2430 &                          & 4 - Optical small amplitude regular variability                & This work \\              
20149+3440 &                          & 4 - Young massive star candidate; questionable OH detection    & This work \\              
20160+2734 & AU\,Vul                  & 2 - Post-AGB star                                              & \cite{Suarez06} \\        
20547+0247 & U\,Equ                   & 2 - Post-AGB star                                              & \cite{Lagadec11} \\       
\noalign{\smallskip}\hline
\end{tabular} 
\end{center}
\end{table*}

Not counting IRAS\,18551+0159 and IRAS\,19200+1035, which had been erroneously reported as OH maser detections (see below), Table \ref{T:no-ohir} lists 16 objects in an evolutionary stage beyond the AGB. These objects having no periodic large-amplitude variations are classified as post-AGB (candidates), (bipolar) proto-planetary nebula (PPN), or planetary nebula (PN) in the literature. In some of them, dubbed as `water fountains', H$_2$O maser emission has been detected tracing bipolar outflows. We refer to all of them with the general term post-AGB star (candidate). They are overplotted on the bulk of the Arecibo sample in the three 2CDs shown in Fig. \ref{F.IRcc}. The location of the post-AGB stars on the 2CDs is strongly dependent on their evolutionary status. After departure from the AGB, post-AGB stars lose their large-amplitude variability and are initially still obscured in the optical and often also in the NIR. This is due to the high opacity provided by the dust in the remnant CSE, which was accumulated at the end of the AGB phase, when the highest mass-loss rates were achieved \citep{Engels02}. Once the CSE starts to disperse, the central star reappears becoming optically bright again. While in the early post-AGB phase the spectral energy distribution (SED) is dominated by the dust of the remnant CSE, the stars in the later phases typically show composite SEDs with two maxima in the optical/NIR and in the mid/far-IR, coming from the star and the CSE, respectively \citep{Kwok93}.

Post-AGB stars of both types are present in the Arecibo sample populating different parts of the 2CDs. In the IRAS 2CD, tracing the flux from the remnant CSE, most of the post-AGB stars accumulate at very red colours ([12]$-$[25]\,$>$\,0.5 mag). Among the few sources with bluer IRAS colours are the optically bright more evolved sources IRAS\,18123+0511 (V2053\,Oph) and IRAS\,20547+0247 (U\,Equ). These optically bright more evolved sources are also conspicuous in the NIR 2CD, where IRAS\,18123+0511 together with IRAS\,07331+0021 (AI\,CMi), IRAS\,18095+2704, (HD\,335675), and IRAS\,20160+2734 (AU\,Vul) show extremely blue NIR colours ($J-H$\,$<$\,0.7), indicating very few dust leftover in the line of sight to the stars. The very red part of the NIR 2CD ($J-H$\,$>$\,3.0 mag) contains no post-AGB stars, because stars in this colour range are usually not detected in all NIR bands. An example is the PN IRAS\,19035+0801 with a colour $H-K$\,=\,4.1 mag and no $J$-band detection.

The separation between AGB and post-AGB stars is apparent also in the near/mid-IR AllWISE 2CD. Except two, the post-AGB stars have very red colours either in the NIR (W1$-$W2\,$>$\,2.0 mag) or in the mid-IR (W3$-$W4\,$>$\,2.0 mag) separating them from the bulk of the LPLAV. The two stars with AllWISE colours matching those of LPLAV are IRAS\,18123+0511 and IRAS\,18577+1047. The post-AGB classification of the latter star is discussed further in Appendix \ref{Ap.notes}.

There are in Table \ref{T:no-ohir} four post-AGB candidates, which were classified by us on the basis of their very red colours and the non-detection of periodic large-amplitude variability. Two were identified already in Paper\,I. More detailed information on these sources is given in Appendix \ref{Ap.notes}.

\subsubsection{The steep brightness increase of IRAS\,19566+3423}

\begin{figure}
\includegraphics[width=\linewidth]{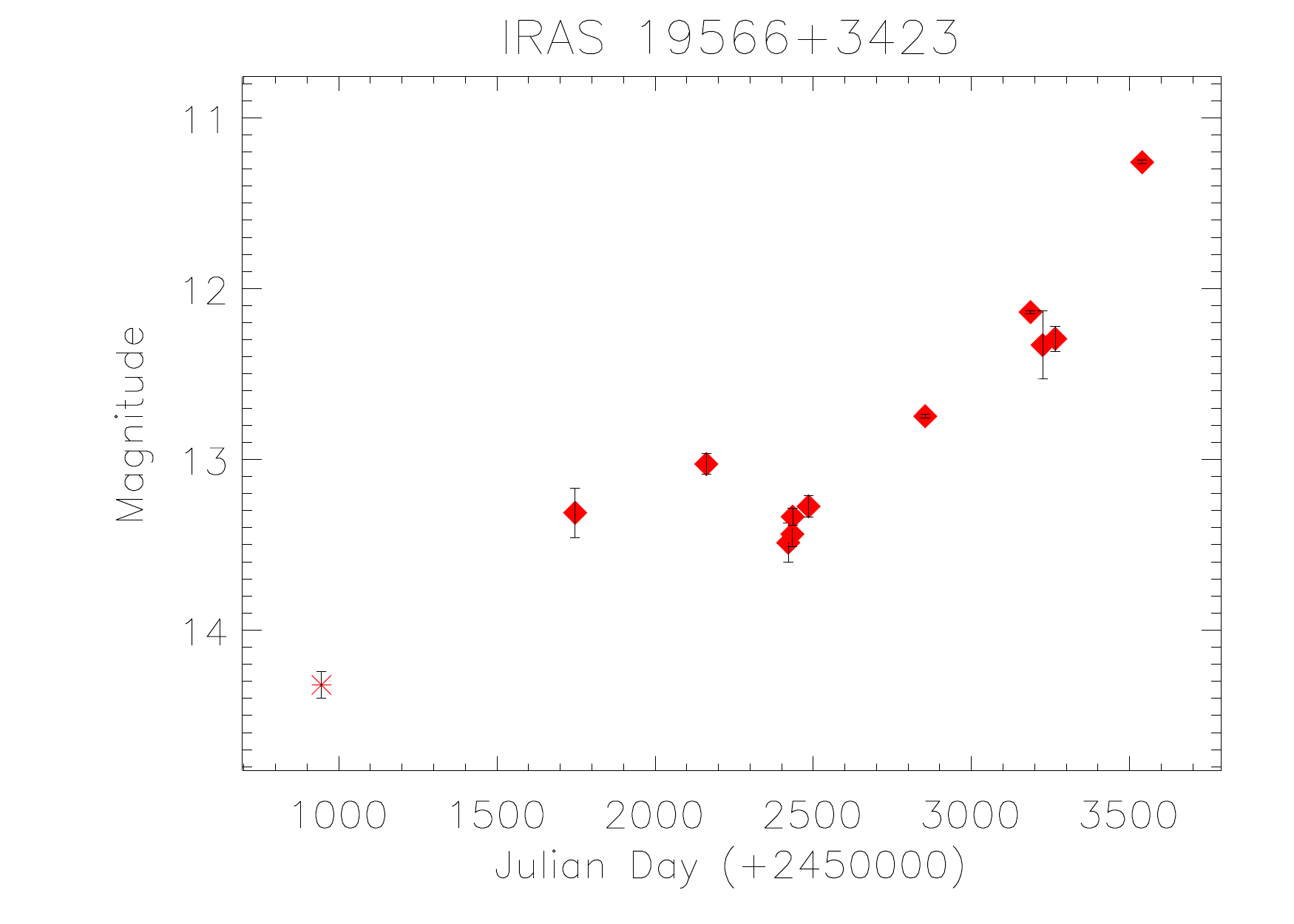}
  \caption{$K$-band light curve from the NIR-DB for IRAS\,19566+3423. The asterisk corresponds to the 2MASS data, and the diamonds to our NIR-MP. Photometric error bars are overplotted.}
  \label{F.19566LC}
\end{figure}

An interesting case is the post-AGB star IRAS\,19566+3423. This source is among the reddest at the far- and mid-IR wavelength range and it showed a steady increase of its $K$-band emission for more than 3 mag between the years 1998 and 2005, while it was not detected in the $J$ and $H$ bands (see Fig.\,\ref{F.19566LC}). \cite{Kamizuka20} studied the $K$-band light curves of a sample of non-variable OH/IR stars finding for several of them a typical brightening rate from 0.01 to 0.13 mag\,yr$^{-1}$. They modelled the increase in the NIR emission as the result of the dilution of the CSE into space and the reappearance of the central star. Their model could not explain the largest brightening rate of 0.33 mag\,yr$^{-1}$ observed for OH\,31.0$-$0.2, and they therefore discussed additional causes for this case. IRAS\,19566+3423 is a new example of an extreme brightening on a time-scale of years with an even larger rate ($\sim 0.43$ mag\,yr$^{-1}$).

\subsection{Unclassified sources}

A small number of sources (IRAS\,05284+1945, IRAS\,18475+0353, IRAS\,18501+0013, IRAS\,18517+0037, IRAS\,19006+0624, IRAS\,19178+1206, IRAS\,19188+1057, IRAS\,19374+1626, and IRAS\,19440+2251), which were not detected by the NIR-MP (cf. Table \ref{t:no-phot}) and have not been studied elsewhere remain unclassified. All of them have red colours and are located in the IRAS 2CD at colours [12]$-$[25]\,$>$\,0.5 mag, where only OH/IR stars with periods P\,$>$\,500 d or post-AGB stars are located (Fig. \ref{F.IRcc}). In the AllWISE 2CD, they have extremely red colours (W1$-$W2\,$>$\,2.3 mag) overlapping with colours of the extreme OH/IR and post-AGB stars. The red colours in both 2CDs are consistent with optically thick CSEs capable to obscure the central star making the object undetectable in the NIR with the instrumentation we used. Thus, they are probably heavily obscured LPLAV AGBs or infrared post-AGB stars in a very early state of their evolutionary phase. A similar conclusion was already drawn in Paper\,I and in \cite{Jimenez-Esteban06b}.

The two other unclassified sources in our sample are IRAS\,19029+0933 (contaminated photometry) and IRAS\,19060+1612 (poorly sampled light curves), which are discussed in Appendix\,\ref{Ap.notes}.

\subsection{Miscellaneous sources}

Due to the original IRAS colour selection of the Arecibo sample, contamination with other types of sources than post-AGB stars is not surprising. Young stellar objects (YSOs) especially are overlapping in colours with the extreme OH/IR stars and they were removed from the sample as soon as they could be identified (\citealt{Lewis97}; Paper\,I). In the current sample, two more YSOs are present: IRAS\,05506+2414 and IRAS\,06319+0415 (Table \ref{T:no-ohir}). Their classification is in line with the absence of large-amplitude periodic variability.

In addition, we identified three RSGs: IRAS\,19083+0851, IRAS\,19229+1708, and IRAS\,20015+3019 (Table \ref{T:no-ohir}). Their colours largely overlap with those of AGB stars \citep{Messineo12}, so that their presence in the sample was expected, although with small numbers because of their rareness. We discuss the available information on these sources individually in Appendix \ref{Ap.notes}.

For four sources with unusual variability properties only preliminary classifications can be given. This is the case of IRAS\,02404+2150, IRAS\,19189+1758, and IRAS\,20127+2430, for which we found periodic small amplitude variations in the optical wavelength range, while we were not able to confirm these with the NIR-DB. Typical optical brightnesses are $G$\,=\,$12-14$ mag, periods $<$\,500 d, and amplitudes $<$\,1 mag. We observed irregular NIR light curves for IRAS\,19500+2239. Details on these sources can be found in Appendix \ref{Ap.notes}. Their colours are consistent with colours of LPLAV, but they are probably semiregular or irregular variable stars.

Finally, four Arecibo sources listed in Table \ref{T:no-ohir} need to be removed from the sample, because the reported OH maser detections are either spurious (IRAS\,06238+0904 and IRAS\,20149+3440) or sidelobe responses (IRAS\,18551+0159 and IRAS\,19200+1035). Also, these cases are discussed in Appendix \ref{Ap.notes}.

\section{Conclusions.}
\label{sec:conclusions}

In this work, we studied the variability properties of the Arecibo sample of OH/IR stars. This sample is composed of 385 sources, and it contains a large portion ($\sim$\,1/3) of heavily obscured AGB stars. To study its variability properties, we conducted a dedicated NIR monitoring program over 6 yr, and we made use of optical and NIR data collected from publicly available surveys and from the literature as complementary information.

We made a fit to the NIR and optical light curves using an asymmetric cosine function, and obtained periods, peak-to-peak amplitudes, and asymmetry factors. Combining the optical with the NIR results, we were able to assign a unique period for 345 out of 385 Arecibo sources. Another three periods were obtained from the literature. The rest of the sources showed semi- or irregular variability or no variability information could be achieved.

Based on the variability properties, we classified the Arecibo sources: 345 ($\sim90$\%) out of 348 objects with periods are LPLAVs on the AGB; 16 are in a later (post-AGB) evolutionary state; 11 remained unclassified, although 9 of them are likely extreme OH/IR stars; and the rest resulted to be different types of sources including 3 objects with small amplitude periodicity.

The period distribution of the Arecibo LPLAVs peaks at $\sim 400$ d, with periods between 300 and 800 d for most of the sources, and it has a long tail up to $\sim 2100$ d. The peak-to-peak amplitudes vary from $\sim 0.5$ up to 5 mag in the NIR, and from $\sim 2$ up to $\sim 7$ mag in the optical, decreasing with wavelength. Most of the sources have light curves with steeper rising branch in comparison with the descending branch. Compared to the sample of O-rich AGB stars compiled by \cite{Suh17}, the periods are on average longer. For $P<900$ d the period distributions of the Arecibo sample and the IRAS selected sample of \cite{Urago20} are compatible. The very long periods ($P>900$ d) present in the Arecibo sample are not in either of the two comparison samples, although extreme OH/IR stars with such long periods have previously been reported in the literature.
 
We confirm the previously known correlation between periods and amplitudes for Mira variables and variable OH/IR stars, in the sense that longer periods are correlated with larger amplitudes. Periods, on average, also increase with larger infrared colours.

One post-AGB star, namely IRAS\,19566+3423, shows a remarkable increase of brightness ($\sim 0.43$ mag\,yr$^{-1}$) over $\sim7$ yr. This increase could be due to the dilution of its CSE on a time-scale of few decades after the source had left the AGB and it is rapidly decreasing its mass-loss rate. However, the rate of brightening surpasses the upper limit determined by \cite{Kamizuka20} in their models for the $K$-band brightening rate. Because of its high brightness (K\,$\sim 11$ mag in 2005), IRAS\,19566+3423 is an ideal target for follow-up observations to study the short AGB to post-AGB transition phase.

The photometric data obtained by our NIR monitoring program, augmented by public surveys, are available online, as well as all light curves and the results of their analysis.

\section*{Acknowledgements}
We wish to thank for the support provided during the observational campaigns by L.~Agudo-M\'erida, C.~Fechner, B.~Fuhrmeister, A.~Manchado, and F. J.~Zickgraf. We thank R. Urago for discussions and providing us with light curves from their 2007$-$2014 monitoring observations at Iriki observatory. We thank the referee for numerous suggestions, which led to an improvement in the presentation of the data.

Support was granted by Deutsche Forschungsgemeinschaft through grant EN~176/30, by the Spanish Mnistero de Ciencia e Innvac\'on (MICINN) under the Consolider-Ingenio 2010 Program grant CSD2006-00070 First Science with the GTC\footnote{\url{http://www.iac.es/consolider-ingenio-gtc}}, by the Spanish MINECO/FEDER through the grants AyA2011-24052 and AyA2017-84089, by the CoSADIE Coordination Action (FP7, Call INFRA-2012-3.3 Research Infrastructures, project 312559), by the Spanish State Research Agency (AEI) Project MDM-2017-0737 at Centro de Astrobiolog\'ia (CSIC-INTA), Unidad de Excelencia Mar\'ia de Maeztu, and by ESCAPE - The European Science Cluster of Astronomy \& Particle Physics ESFRI Research Infrastructures, which received funding from the European Union's Horizon 2020 research and innovation programme under grant agreement no. 824064. D.A. thanks the Leverhulme Trust for financial support. 

This publication made use of the infrared data products from: the Wide-field Infrared Survey Explorer, UKIRT Infrared Deep Sky Survey, Two Micron All Sky Survey, and DENIS. This work is also partially based on optical data from: ASAS and ASAS-SN, the OMC Archive at CAB (INTA-CSIC), the European Space Agency (ESA) mission {\it Gaia}, and the AAVSO International Database contributed by observers worldwide. We made intensive use of {\sc IRAF}, {\sc Astrometry.net}, and {\sc Topcat} \citep{Taylor05}. This research has also made use of the VizieR catalogue access tool, CDS, Strasbourg, France. Finally, we acknowledge the use of the ADS bibliographic services.



\section*{Data Availability}
\label{sec:online}
The data underlying this article are online available in the article at MNRAS and in its online supplementary material: the full electronic versions of the Tables \ref{t:sample}, \ref{t:phot}, \ref{t:photosummary}, and \ref{t:fitsummary}, and figures in Appendices \ref{Ap.LC} and \ref{Ap.Fits}. Tables are also available at CDS via anonymous ftp to cdsarc.u-strasbg.fr (130.79.128.5) or via \url{http://cdsarc.u-strasbg.fr/viz-bin/cat/J/MNRAS/505/6051}.


\bibliographystyle{mnras}
\bibliography{references} 




\appendix

\section{Notes on individual sources.}
\label{Ap.notes}
In this section, we discuss individual sources, with emphasis on their variability properties. We refer the reader to Appendix\,\ref{Ap.NIR-LC} for the NIR light curves, and to Appendix\,\ref{Ap.Opt-LC} for the optical ones.

\begin{itemize}
    \item IRAS\,02404+2150: This source presents a well-sampled light curve in the optical, with a clear regular variability but with a small amplitude ($\sim 0.8$ mag) in comparison with LPALVs (cf. Fig. \ref{F.Amplitudes}). The optical period of $\sim 380$ d could not be reproduced from the NIR light-curves fits, possibly because of sparse sampling and even smaller amplitudes than tose in the optical. The variability characteristics of the star were discussed by \cite{Whitelock95}, who classified it as a low-amplitude semiregular variable. They argue that the redness of the star could be due to dust accumulated during a previous Mira phase, and the current low-amplitude variability would be a consequence of a recent helium flash leading to the loss of large-amplitude pulsations. Alternatively, the redness could be due to some extra dust created by interactions with a companion within a binary system.
\item IRAS\,06238+0904 (RAFGL 940) is a carbon star, for which OH maser emission would not be expected. \cite{Chengalur93} detected a single 29\,mJy per beam faint line at 1612 MHz. This maser line was never reobserved, and at other transitions of OH \citep{Lewis97}, or H$_2$O at 22 GHz (\citealt{Engels96}; \citealt{Yung13}) no maser emission had been detected so far. We suspect therefore that the original detection was spurious.
    \item IRAS\,18520+0533: The source was discussed in Paper\,I as post-AGB candidate because of its red IRAS colour [12]$-$[25]\,=\,1.75 connected to a relatively bright NIR counterpart. The classification is corroborated here by the verification of its non-variability.
\item IRAS\,18551+0159: This was a false OH maser detection based on the coincidence of the reported velocities of the OH maser features with the nearby source IRAS\,18549+0208 \citep{Engels15b}. IRAS\,18551+0159 is considered as planetary nebula candidate GPSR\,035.472$-$0.436 \citep{Zhu13}.
    \item IRAS\,18577+1047: The star was considered by \cite{Lewis89} as post-AGB candidate based on the relative strength of its maser emission in different transitions, in particular of its relatively strong emission in the OH main lines. The small-amplitude variability supports this classification. However, its infrared colours ($J-H$\,=\,1.31, $H-K$\,=\,0.75, W1$-$W2\,=\,1.08, W3$-$W4\,=\,1.71, [12]$-$[25]\,=\,$-$0.08, [25]$-$[60]\,=\,$-$1.30) are indistinguishable from the colours of ordinary LPLAVs (Fig. \ref{F.IRcc})
\item IRAS\,19029+0933: The NIR photometry of the source is blended with a nearby field star. The brightness of both objects is similar in the NIR, while in the mid- and far-IR IRAS\,19029+0933 dominates the emission. Its colours W1$-$W2\,=\,1.13, W3$-$W4\,=\,1.28, [12]$-$[25]\,=\,$-$0.16, [25]$-$[60]\,=\,$-$1.34 are not particularly red, and would be consistent with a classification as (LPLAV) AGB star. 
    \item IRAS\,19060+1612: It is a relatively bright star (G\,$\sim 11$; K\,$\sim 4$ mag) close to the Galactic plane ($b$\,=\,+3.7\degr). The optical light curve shows irregular variability with small amplitude ($< 0.8$ mag). The variability is also detected in the NIR, but the small number of observations prevents a closer analysis of the light curves. The original 1612 MHz OH maser detection \citep{Chengalur93} relied on a single line with marginal signal-to-noise ratio and it may be spurious. Additionally, it was not detected in the main OH maser lines at 1665 and 1667\,MHz \citep{Lewis97}, nor in the H$_2$O maser transition \citep{Engels96}. Therefore, we are not able to confirm this object as an AGB star.
\item IRAS\,19065+0832 (OH\,42.6+0.0): Similarly to IRAS\,18520+0353, the source has a relatively bright NIR counterpart compared to its red [12]$-$[25]\,=\,1.02 colour. Its non-variability corroborates the classification as post-AGB candidate.
    \item IRAS\,19083+0851: The object was classified as RSG by \cite{Engels96} on the basis of its large OH maser velocity range $\Delta v_{\rm OH}$\,=\,63 km\,s$^{-1}$ \citep{Lewis94}. The star is optically faint (G\,$>$\,16 mag). Its optical light curve from \emph{Gaia} shows variability but it contains only a dozen of measurements in less than 2 yr, which prevents any further analysis. At the NIR bands, it shows only small-amplitude variations and we found no periodicity. IRAS\,19083+0851 is located in the direction of the W49 star-forming complex and is not considered a YSO, but classified as an AGB candidate instead \citep{Saral15}. Based on the maser properties of this star and the absence of large amplitude variations, we consider the AGB classification as not justified, however.
\item IRAS\,19177+1333: The NIR flux is constant within the errors. The source is also located in the direction of the W49 star-forming complex and could be a young star in agreement with the 1612 MHz OH maser spectrum consisting of only a single line. However the maser line is weak ($\sim50$ mJy; \citealt{Lewis90b}), so that the secondary peak of a double-peaked profile indicating an evolved star may be weaker and may escaped detection. IRAS\,19177+1333 was not detected in other maser transitions at 22 GHz H$_2$O \citep{Engels96}, at 1667 and 1665\,MHz main-line OH \citep{Lewis97}, or at 43 GHz SiO \citep{Nakashima03}. \cite{Saral15} excluded this source as YSO; therefore, we consider the source as post-AGB candidate, because of its non-variability and its very red [12]$-$[25]\,=\,0.74 mag colour.
    \item IRAS\,19189+1758: The optical light curve of this source was obtained from sparsely sampled \emph{Gaia} data with 24 observations. We derived a reliable period of $\sim 440$ d, which is in perfect agreement with the period P\,=\,434 d listed in the ATLAS Catalogue \citep{Heinze18}. The period P\,=\,218 d derived by \cite{Wozniak04} from NSVS data must be considered as an alias. However, the optical amplitude ($\sim 0.7$ mag) is rather low as to be considered as typical for an LPLAV. The NIR light curves are among the best sampled but the brightness variations are small ($0.5-0.8$ mag; Table\,\ref{t:photosummary}). We could not obtain good fit solutions in none of the three bands. Because of its blue [12]$-$[25]\,=\,$-0.49$ mag colour, this source is likely a semiregular variable of type SRa. This source may be in a similar evolutionary state than IRAS\,02404+2150.
\item IRAS\,19200+1035: This object is the known planetary nebula K\,3$-$33 \citep{Acker92}. \cite{Engels15b} found that the reported OH maser detection was, however, just due to confusion with the nearby source IRAS\,19201+1040.
    \item IRAS\,19229+1708: The star was classified by \cite{Winfrey94} as M-Supergiant of luminosity class M3-4\,I based on optical spectra. The classification is supported by a relatively large expansion velocity of the CSE $v_{\rm exp}$\,$\sim20$ km\,s$^{-1}$ \citep{Engels96}. Its optical $G$-band brightness is $G$\,$\sim13$ mag. Similarly to IRAS\,19083+0851, the optical \emph{Gaia} light curve shows variability with small amplitude ($\sim1$ mag). The baseline ($<$\,2 yr) is, however, too short to derive further conclusions on its variability properties. In addition, only small-amplitude variations were observed in the NIR bands. In summary, all available observations are consistent with the RSG classification.
\item IRAS\,19319+2214: The source appeared already in samples of post-AGB and PN candidates (\citealt{Ramos-Larios12}; \citealt{Gomez15}) because of its red IRAS colours. The OH maser at 1612\,MHz was discovered by \cite{Lewis90b} and showed a peculiar OH maser profile different from those usually seen in OH/IR stars. It consists of a well-defined pair of outer peaks, as well as an additional pair of inner features symmetrically placed relative to the central velocity. This pattern suggests the presence of two shells in the CSE \citep{Lewis04}. H$_2$O maser emission was detected by \cite{Engels96} within the velocity range of the OH maser. A search for H$_2$O masers at velocities outside the OH velocity range was unsuccessful \citep{Gomez15}. NIR observations from 2008 by \cite{Ramos-Larios12} are consistent with the absence of variations compared to our observations ($1999-2005$), which showed a small brightness increase after $\sim2002$ (JD\,$\sim$\,2452500). Maser and IR observations are therefore consistent with a classification as a post-AGB star.
    \item IRAS\,19500+2239: The NIR light curves show clear large-amplitude variability, but we could not determine any periodicity. Parallel to the NIR-MP, the star was monitored in the $K$ band in the period $2000-2005$ by \cite{Tang08}, who also observed strong irregular variations. The IR colours are similar to the colours of LPLAV, so IRAS\,19500+2239 may be an AGB star, but with peculiar variability characteristics. 
\item IRAS\,20015+3019 (V719\,Cyg): The star was classified by \cite{Winfrey94} as M-Supergiant of luminosity class M4\,I on the basis of optical spectra. The spectral classification was confirmed by \cite{Tang08}, who determined a type later than M5 from a new optical spectrum. The maser properties are inconspicuous with single line detections for 1612 MHz OH, H$_2$O, and SiO (\citealt{Chengalur93}; \citealt{Engels96}; \citealt{Kim13}). The CO(1$-$0) emission observed by \cite{Margulis90} extends between 7 and 25 km\,s$^{-1}$. It suggests a wind outflow velocity of 9 km\,s$^{-1}$, which is more typical for AGB stars than RSGs. The NIR-DB provides too few observations to allow a proper classification of the $J$, $H$, and $K$ light curves, but the photometry is compatible with a better sampled $K$-band light curve obtained by \cite{Tang08}. These observations taken in 2000-2005 give an irregular behaviour of the light curve with peak-to-peak amplitudes $\la$1 mag. A periodic light curve with P\,=\,1019 d and an amplitude of $\Delta R_{KELT} \sim1$ mag was observed in the KELT Survey \citep{Arnold20}. Its optical (ASAS-SN and \emph{Gaia}) light curve shows a peculiar shape with a periodicity compatible with the period obtained from the KELT survey. The observations available exclude a classification of IRAS\,20015+3019 as LPLAV. The small-amplitude variability in the NIR is compatible with an RSG classification, but a classification as a semiregular variable AGB star cannot be excluded. For a distance of 2.3 kpc \citep{Bailer-Jones20} we determined a luminosity of $\sim$\,35,000 \Lsun, using VOSA (see Section \ref{var-cycle}). This corresponds to $M_{bol} \approx$\,$-$6.6 mag, which is within the luminosity range of known RSG \citep{Messineo19}.
    \item IRAS\,20127+2430: This star has variability characteristics similar to IRAS\,19189+1758. The optical light curve was obtained from sparsely sampled \emph{Gaia} data with 36 observations. We derived a reliable period of $\sim400$ d, but with a rather small amplitude ($\sim0.3$ mag). The NIR light curves are well sampled but the brightness variations are small ($0.4-0.5$ mag; Table\,\ref{t:photosummary}), in accordance with the small variations in the optical. Like in IRAS\,19189+1758, we could not obtain good fit solutions in none of the three NIR bands. The star may be in a similar evolutionary state than IRAS\,02404+2150.
\item IRAS\,20149+3440: It is the source with the reddest [25]$-$[60]\,=\,3.12 IRAS colour. It is blended in the NIR with another source and shows small variability. The source is in the line of sight of an interstellar dark cloud (DOBASHI\,2265; \citealt{Dobashi11}) with a reflection nebula (GN\,20.14.9) \citep{Magakian03} at less than 8\arcsec. We suspect that the very red IRAS colour is in fact due to contamination by the thermal emission of the dark cloud, specially at 60\,\mic. So, we suggest that IRAS\,20149+3440 is a young massive star. The weak OH maser emission from the original detection could not be confirmed by follow-up observations with the Arecibo radio telescope (B.M. Lewis, private communication); therefore, we consider the original detection as spurious. 
\end{itemize}

\section{Light Curves}
\label{Ap.LC}

Light curves are available as online material.

\subsection{Near-infrared Light Curves}
\label{Ap.NIR-LC}

Here, we present the NIR light curves of the Arecibo sample. We use a colour code for the NIR bands: $J$ in blue, $H$ in green, and $K$ in red. Additionally, different symbols are used depending on the origin of the data: triangles for the NIR-MP $J$ band, squares for the NIR-MP $H$ band, diamonds for the NIR-MP $K$ band, asterisks for the 2MASS data, crosses for DENIS, and pluses for UKIDSS. Photometric errors are overplotted with black vertical bars. They are typically smaller than the symbols.

\subsection{Optical Light Curves}
\label{Ap.Opt-LC}

Here, we present the optical light curves of the Arecibo sample. For the \emph{Gaia} data, we used the $G_{BP}$ photometry to combine with data from other surveys, and the \emph{Gaia} $G$ band when the light curve contains only the \emph{Gaia} data. The code of symbols is as follows: red diamonds for data from ASAS, black crosses for AAVSO, cyan asterisks for the OMC, blue triangles for ASAS-SN, and green squares for \emph{Gaia}. Photometric errors are shown with vertical black bars. They are typically smaller than the symbols.

\section{Results of the light-curve fit}
\label{Ap.Fits}

Here, we present the results from the fits to the observed light curves, which passed all quality controls (see Section \ref{sec:LC-fit}). For each light curve, three figures are available as online material:
\begin{itemize}
    \item The periodogram: It shows the $\chi^2$ as a function of period. We used these diagrams to identify the period that better fits the observational data, as well as other alias periods. The period adopted in this work is labelled at the top of the plot.
    \item The light curve: It shows the observational data and the light-curve model that was fitted to them.
    \item The folded light curve: It shows the observational data distributed along two variability cycles, together with the light-curve model fitted.
\end{itemize}
Symbols are as in Appendices\,\ref{Ap.NIR-LC} and \ref{Ap.Opt-LC}. The model light curves are shown with black solid lines.

\bsp	
\label{lastpage}
\end{document}